\documentclass{aa}  

\usepackage{graphicx}
\usepackage{txfonts}
\usepackage{lipsum}
\usepackage{subcaption}         % necessary for continued figures, example in section 3
\usepackage{lscape}             % to rotate a single page table, example in appendix.
\usepackage{placeins}           % useful with \FloatBarrier, to keep 
\usepackage{hyperref}
\hypersetup{
    colorlinks=true,
    linkcolor=blue,
    filecolor=magenta,      
    urlcolor=cyan,
    pdftitle={Overleaf Example},
    pdfpagemode=FullScreen,
    linkcolor=blue,  % color of internal links (change box color with  linkbordercolor)
    citecolor=blue,        % color of links to bibliography
}
\usepackage{graphicx}	% Including figure files
\usepackage{amsmath}	% Advanced maths commands
\usepackage{soul} % Highlighting

\graphicspath{{./figs/}}

\usepackage{amsfonts}
\usepackage{mathtools}
\usepackage[hypcap=false]{caption}
\usepackage{csquotes}
\usepackage{subcaption}
\usepackage{bm}
\usepackage{bbm}
\usepackage[table,xcdraw]{xcolor}
\usepackage{placeins}
\usepackage{url}
\usepackage{arydshln}
\usepackage{booktabs}
\usepackage{lipsum}
\usepackage{orcidlink}

\usepackage{pifont}

\usepackage[ruled,linesnumbered, vlined]{algorithm2e}

\SetKwRepeat{Do}{do}{while}

\makeatletter
\newenvironment{dataavailability}{%
  \par\tiny\vspace{2ex}\noindent
  {\itshape Data Availability.}\enspace\ignorespaces
}{\par\vspace{1ex}}
\makeatother
\let\orgautoref\autoref
\providecommand{\Autoref}[1]{\def\equationautorefname{Equation}
\def\figureautorefname{Figure}\def\sectionautorefname{Section}\def\subsectionautorefname{Section}\def\subsubsectionautorefname{Section}\def\tableautorefname{Table}\def\algorithmautorefname{Algorithm}\orgautoref{#1}}
\renewcommand{\autoref}[1]{\def\equationautorefname{Eq.}
\def\figureautorefname{Fig.}\def\sectionautorefname{Sect.}\def\subsectionautorefname{Sect.}\def\subsubsectionautorefname{Sect.}\def\tableautorefname{Table}\def\algorithmautorefname{Algorithm}\orgautoref{#1}}

\begin{document}

%%%%%%%%%%%%%%%%%%%%%%%%%%%%%%%%%%%%%%%%
% if you use custom commands in your title,
% ensure to check your title when submitting!
%%%%%%%%%%%%%%%%%%%%%%%%%%%%%%%%%%%%%%%%

   \title{Point spread function wavefront recovery from in-focus stellar observations}
   \subtitle{}
   \titlerunning{PSF wavefront recovery}
   \authorrunning{E. Centofanti et al.}

%%%%%%%%%%%%%%%%%%%%%%%%%%%%%%%%%%%%%%%%
% Please separate each author with the \and command
%
% Please do not include ORCIDs next to author names.
% Only ORCIDs authenticated by individual authors in EDPS
% editorial system will be taken into account.
% ORCIDs included here will be removed.
%%%%%%%%%%%%%%%%%%%%%%%%%%%%%%%%%%%%%%%%

    \author{
    Ezequiel Centofanti\inst{1}\fnmsep\thanks{E-mail: \href{mailto:ezequiel.centofanti@cea.fr}{ezequiel.centofanti@cea.fr}}\orcidlink{0009-0001-8461-8451}
    \and Samuel Farrens\inst{1}\orcidlink{0000-0002-9594-9387}
    \and Jean-Luc Starck\inst{1,3}\orcidlink{0000-0003-2177-7794}
    \and Tob{\'i}as I.~Liaudat\inst{2}\fnmsep\thanks{E-mail: \href{mailto:tobias.liaudat@cea.fr}{tobias.liaudat@cea.fr}}\orcidlink{0000-0002-9104-314X}
        }

   \institute{
    AIM, CEA, CNRS, Université Paris-Saclay, Université de Paris, F-91191 Gif-sur-Yvette, France
    \and IRFU, CEA, Universit{\'e} Paris-Saclay F-91191 Gif-sur-Yvette Cedex, France
    \and Institutes of Computer Science and Astrophysics, Foundation for Research and Technology Hellas (FORTH), GR-70013 Heraklion, Greece
    }

   \date{Received December, 2025}

  %%%%%%%%%%% Abstract %%%%%%%%%%%
  \abstract{
  Recovering the wavefront error (WFE) field of an optical system from intensity in-focus observations is a challenging inverse problem with broad implications for telescope point spread function (PSF) modelling. Accurate WFE recovery enables both precise PSF modelling and direct insight into the state of the telescope optics, facilitating the detection of potential malfunctions.
  Recently, non-parametric PSF models have shown promising performance in modelling complex optical systems in space-based telescopes. WaveDiff is a semi-parametric PSF model that represents the PSF in wavefront space by combining parametric and learnable features with a differentiable forward optical model. This parameterisation enables phase retrieval from in-focus observations by exploiting the spatial variation of the PSF across the field of view (FOV).
  The original version of WaveDiff achieves outstanding PSF recovery results in pixel space; however, the recovered WFE is far from the ground truth, with a relative error of around $30\%$. In this paper, we present a new optimisation scenario that bridges WaveDiff’s parametric and non-parametric components through wavefront feature projection, yielding a substantial improvement in WFE recovery and making WaveDiff the first demonstrated method to combine wide-field WFE recovery, in-focus-only polychromatic observations, and non-parametric wavefront features in a single framework.
  We show that incorporating wavefront projections and increasing the number of optimisation cycles enables WaveDiff to recover the WFE with an error of approximately $3\%$ using only noisy, undersampled, in-focus observations. This represents a tenfold improvement over the original model while further reducing the pixel-space error. The code to reproduce the results of this article is publicly available at \href{https://github.com/tobias-liaudat/wf-psf/tree/v1.4.0}{https://github.com/tobias-liaudat/wf-psf/tree/v1.4.0}.
  }
%   % context heading (optional)
%   % {} leave it empty if necessary  
%    {Optional, leave empty if necessary.  The heading “Context” is used when needed to
% give background information on the research conducted in the paper}
%   % aims heading (mandatory)
%    {Mandatory. The objectives of the paper are defined here.} 
%   % methods heading (mandatory)
%    {Mandatory. The methods of the investigation are outlined here}
%   % results heading (mandatory)
%    {Mandatory. The results are summarized here.}
%   % conclusions heading (optional), leave it empty if necessary
%    {Optional, leave empty if necessary.  “Conclusions” can be used to
% explicit the general conclusions that can be drawn from the paper.}


   \maketitle
   \nolinenumbers

%=======================================================================================
\section{Introduction}
%=======================================================================================
In the era of precision cosmology, current and upcoming optical telescopes such as \textit{Euclid} \citep{euclid2024, laureijs2011}, the Vera C. Rubin Observatory \citep{Ivezi__2019, LSST2009}, Nancy Grace \textit{Roman} \citep{akeson2019widefieldinfraredsurvey, wfirst} and Xuntian, also known as the Chinese Space Station Telescope (CSST), \citep{csst2025, zhan2011} will provide unprecedented sky coverage, depth, and spatial resolution. 
Accurate modelling of their instrumental response is therefore key to obtaining unbiased and competitive cosmological constraints. 
The instrumental response, commonly referred to as the point spread function (PSF), describes how incident light is blurred as it passes through the optical system of the telescope. 
Thanks to the depth and wide coverage of modern space surveys, these facilities will enable detailed studies of the large-scale structure of the Universe through statistical probes, such as weak gravitational lensing (WL) \citep{kilbinger2015, mandelbaum2018_bis3}. 
Measuring the WL signal requires extremely precise estimates of galaxy shapes for which the PSF of the instrument plays a fundamental role and is a dominant source of systematic uncertainty in the era of precision cosmology \citep{massey2012}.

Accurately modelling the PSF of wide-field space telescopes that operate in a single, very broad optical band (such as the \textit{Euclid} VIS instrument \citep{cropper2016}) is extremely challenging. 
The PSF varies spatially across the field of view (FOV), as a function of wavelength, and also over time, primarily due to temperature fluctuations in the components of the optical system of the telescope. 
There are two main approaches to PSF modelling \citep[for a review see e.g.][]{liaudat2023_2}: parametric and non-parametric. 
The parametric approach relies on a physical model of the optical system of the telescope, where a small set of parameters needs to be calibrated by combining technical measurements of the instrument and on-sky observations. 
Given the demanding performance requirements, this pre-flight calibration is often insufficient, creating a calibration gap that must be compensated using non-parametric methods which offer greater flexibility as they do not attempt to model the underlying physics of the optical system.
A parametric approach was initially used for the Hubble Space Telescope (HST) with the Tiny Tim model \citep{krist2011}, which was later improved upon by a non-parametric method \citep{HST_PSF2017, gillis2020validation}.
Non-parametric PSF modelling circumvents the need for technical measurements by directly modelling the PSF from observations. 
Non-parametric PSF models, such as PSFEx \citep{bertin2011}, RCA \citep{ngole2016}, MCCD \citep{liaudat2020}, \textit{lensfit} \citep{miller2007, kitching2008, miller2013}, among others, base their representation of the PSF field on a decomposition of PSF features in pixel-space learned from observations of distant stars. 

WaveDiff \citep{liaudat2023} is a novel semi-parametric PSF modelling approach that represents the PSF in wavefront space, and fit it using in-focus stars only. 
This wavefront error (WFE) representation is propagated through a fully differentiable model of the optical system of the telescope. 
By modelling the PSF in wavefront space, WaveDiff is able to model the spectral variation of the PSF due to diffraction across the entire FOV, which, to the best of our knowledge, no other non-parametric model is currently capable of doing. 
This is particularly relevant for WL studies, as the PSF is derived from stellar observations but applied to galaxies with different spectral energy distributions (SEDs). 
The telescope WFE estimation not only allows for the construction of a very precise PSF model but also enables the diagnosis of potential issues in the optical system. 
For example, the James Webb Space Telescope (JWST) \citep{gardner2006james} uses wavefront sensing to detect micrometeorite impacts on its primary mirrors \citep{JWST_Performance_2023}. 

In \citet{liaudat2023} the authors introduce the WaveDiff model and present PSF field reconstruction results both in pixel and wavefront space. 
Their semi-parametric model is capable of representing the PSF field in pixel space with a relative error of less than $1\%$. 
However, since the non-parametric wavefront features were directly learned from the data, the estimated WFE representation does not coincide with the ground truth, yielding a relative error of around $100\%$. 
These results illustrate the difficulty of estimating the WFE from low-resolution, in-focus observations. Typically, WFE estimation benefits from phase-diverse, out-of-focus observations \citep{fienup1993phase, fienup1993} in order to break degeneracies.
However, defocusing the optical system is not always feasible, as it can pose risks to space telescopes and results in a loss of observation time, as the instrument needs to stabilise once it is back in focus.
%Work such as \cite{Jurling_2012} reconstructs the WFE from multiple in-focus observations using a parametric PSF model built from prior knowledge of the optical design of the instrument.
\cite{Jurling_2012} is the first work to propose exploiting the spatial diversity of the PSF to estimate the WFE from in-focus observations. WaveDiff extends this work by employing a semi-parametric modelling of the WFE across the entire FOV.

In this paper, we present a new optimisation scenario for the WaveDiff semi-parametric PSF model that bridges its parametric and non-parametric components through wavefront feature projection, better exploiting the spatial diversity of the PSF and allowing us to constrain the underlying WFE field from a set of observed in-focus stars.
This scenario is different from standard phase retrieval problems, which typically rely on multiple out-of-focus observations of the same star. 
In our case, each star provides a single in-focus observation and has a different WFE representation. 
However, all of the representations originate from the same underlying WFE field, subject to regularity conditions stemming from the optical system modelling.

This paper is organised as follows. 
In \autoref{sec:psf_modelling}, we introduce the PSF modelling problem for space telescopes; in \autoref{sec:sota}, we discuss the state of the art in phase retrieval for PSF modelling; in \autoref{sec:wavediff}, we present WaveDiff, the PSF model under study in this work; 
in \autoref{sec:methodology}, we introduce the new WaveDiff optimisation scenario and describe the changes with respect to the original optimisation process; \autoref{sec:results} presents the experimental results; 
and finally, we outline future perspectives and conclude this study in \autoref{sec:conclusion}.

\vspace{-.3cm}
%=======================================================================================
\section{The PSF field and the PSF modelling inverse problem for space telescopes}
\label{sec:psf_modelling}
%=======================================================================================
The PSF characterises the effects that the optical system introduces to the image of an observed object. 
When these imaging artefacts can be described as linear, the PSF may be modelled as a spatially varying convolution kernel, that is, a continuous function $\mathcal{H}_{\text{int}}: \mathbb{R}^{2} \times \mathbb{R}_{+} \times \mathbb{R}_{+} \times \mathbb{R}^{2} \to \mathbb{R}_{+}$, which represents the instantaneous monochromatic PSF of the optical system at every position in the FOV. 
Accordingly, we write $\mathcal{H}_{\text{int}}(u,v;\lambda;t|u_i,v_i)$ for the PSF centred at field position $(u_i, v_i)$, wavelength $\lambda$ and a time $t$. Here, $(u,v)$ denote the kernel coordinates of the PSF itself, while the conditioning $(u_i, v_i)$ specify the PSF centre location in the FOV. 
In this paper we follow the notation introduced in \citet{liaudat2023_2}, summarized in \autoref{tb:variable} in the Appendix.
We assume that the PSF field is locally spatially invariant \citep{liaudat2023_2}, such that the PSF field variations on the scale of the objects we are imaging, stars and galaxies, can be neglected. 
Therefore, the observed image produced by the optical system can be expressed as the convolution of the ground-truth object with the PSF,
\vspace{-.2cm}
\begin{equation}
    \mathcal{I}_{\text{obs}}(u,v;\lambda|u_i,v_i) = (\mathcal{H}_{\text{int}} \ast \mathcal{I}_{\text{GT}})(u,v;\lambda|u_i,v_i),
\end{equation}
where we have dropped the time dependency $t$ since we are not considering transient objects, we study exposures independently, and we assume that the PSF field time dependency is much slower than the exposure time. 
$\mathcal{I}_{\text{GT}}$ is the ground-truth object at FOV position $(u_i,v_i)$, $\mathcal{H}_{\text{int}}$ is the intensity impulse response or PSF of the optical system, and $\mathcal{I}_{\rm obs}$ is the observed object. 
The object under study suffers further degradation on the way to obtaining the actual image. 
First, in the case of single-band telescopes, the wavelength-dependent observation is integrated over the passband of the telescope. 
Then, the observation is sub-sampled in the detector \citep{high2007, lauer1999}, discretising the image coordinates $(u,v)$ into pixel indices $(\bar{u},\bar{v})$.
Finally, the sampled image contains observational noise, including thermal noise \citep{nyquist1928}, readout noise \citep{basden2004}, dark-current shot noise \citep{baer2006}, and photon shot noise.
There are other detector effects such as the brighter-fatter effect \citep{coulton2018, guyonnet2015} or charge transfer inefficiency \citep{rhodes2010} that further degrade the image and make the observation model more complex. 
For simplicity, in this article we assume that the observations have been calibrated and corrected for these detector effects and assume a white Gaussian model for the observational noise. 
In this case, the observation model is written as
\vspace{-.3cm}
\small
\begin{multline}
    I_{\text{img}}(\bar{u},\bar{v}|u_i,v_i) = \\
        \mathcal{F}_p \left\{ \int_0^{+\infty} \mathcal{T}(\lambda)\; 
        (\mathcal{I}_{\text{GT}} \ast \mathcal{H}_{\text{int}})
        (u,v;\lambda | u_i, v_i) \; d\lambda \right\} 
        +  N(\bar{u},\bar{v}|u_i,v_i),
        \label{eq:obs_model}
\end{multline}
\normalsize
where $\mathcal{T}(\lambda)$ is the transmission function of the telescope, $\mathcal{F}_p$ represents the pixelisation function, and $N$ is the observational noise. 
Note that for a given FOV position $(u_i, v_i)$ we are considering a single PSF $\mathcal{H}_{\text{int}}(u,v;\lambda | u_i, v_i)$ following the locally spatially invariant approximation. 

When extracting scientific information from astronomical images, as in the case of weak gravitational lensing cosmology \citep{mandelbaum2018_bis3}, the accuracy and precision of the observations are critical.
Consequently, it is necessary to account for the various dependencies of the PSF model in order to meet the stringent requirements of modern surveys.
As shown in \autoref{eq:obs_model}, the PSF depends both on the spatial position within the FOV and on the wavelength.
In addition, the PSF varies over time, which must also be properly modelled. 
This temporal dependence arises from changes in the telescope, such as mechanical deformations in the optical system caused by temperature gradients. 
A common approach to account for temporal variations is to construct a separate PSF model for each exposure, which greatly reduces the number of stars available to estimate the PSF field. 
As in \citet{liaudat2023}, in this paper we neglect the temporal dependence of the PSF.
The spatial variations arise from the different optical paths taken by light entering the system at varying angles of incidence, which consequently accumulate distinct optical aberrations. 
Imaging systems with large focal planes, or wide field-of-view instruments \citep{euclid2024, LSST2009, gwyn2019}, are more susceptible to exhibit stronger PSF spatial variations.
In diffraction-limited systems, spectral variations are primarily influenced by diffraction diffraction phenomena, which introduces a well-studied wavelength dependence. 
The presence of refractive elements as well as of complex dielectric coatings \citep{baron2022} in the optical system can also generate additional spectral dependencies. 
Accounting for these variations is essential because stars are used to constrain the PSF model, whereas the estimated PSF is later applied to extended objects such as galaxies. 
Since stars and galaxies have different spectral energy distributions, a given WFE results in different polychromatic PSFs. 
To address this, it is necessary to accurately estimate the monochromatic PSF, that is, the PSF at each wavelength.

Building a PSF model is a challenging task, not only because of the level of accuracy demanded by modern surveys, but also because we only observe its effect on astronomical objects rather than the PSF itself. 
Moreover, as previously noted, observations are subject to several degradation processes, such as spectral integration, undersampling, and noise, which make an analytical solution intractable. 
Under these conditions, PSF estimation must be treated as an inverse problem.
A case of particular interest for PSF fitting is the observation of unresolved stars. 
These objects can be approximated as spatial impulses, expressed as
\vspace{-.1cm}
\begin{equation}
    \mathcal{I}_\mathrm{star}(u,v;\lambda|u_i,v_i) = \delta(u-u_i,v-v_i) f_{\rm SED}(\lambda),
\end{equation}
where $f_{\rm SED}(\lambda)$ is the spectral energy distribution of the observed star. 
Since the convolution of a function with a Dirac delta reproduces the function shifted to the location of the delta, the observation of a star provides a sample of the PSF at that position. 
However, it is not a direct sample, as imaging degradation must be taken into account. 
% The pixelised observation model for a star is written as follows:
% %
% \begin{multline}
%     I_{\text{star}}(\bar{u},\bar{v}|u_i,v_i) = \\
%         \mathcal{F}_p \left\{ \int_0^{+\infty} \mathcal{T}(\lambda)\, f_{\rm SED}(\lambda)\,
%         \mathcal{H}_{\text{int}}
%         (u,v;\lambda | u_i, v_i) \; d\lambda \right\} 
%         \;+ \\
%         \; N(\bar{u},\bar{v}|u_i,v_i).
%         \label{eq:obs_star}
% \end{multline}
This close relationship makes stellar observations an ideal candidate for PSF model estimation. 

%=======================================================================================
\section{Related works}
\label{sec:sota}
%=======================================================================================

Several PSF models have been developed with the advent of wide-field imagers and weak gravitational lensing studies. 
Most of these models are based on a non-parametric framework to model the spatial variation of the PSF \citep{bertin2011, miller2007, kitching2008, miller2013, ngole2016, liaudat2020}. 
However, these methods do not account for the chromatic variations of the PSF, which is required in broad-band imagers like \textit{Euclid}, LSST or Nancy Grace Roman space telescope. 
%Methods such as \citet{Jurling_2012} and \citet{Fienup_99} take into account the chromatic variation of the PSF using multi-band observations, but they are parametric models that model the PSF based on a few Zernike polynomials.
\citet{Fienup_99} models the spectral variation of the PSF in wavefront space, for individual pointings using defocused star images. \citet{Jurling_2012} is the first work to propose a parametric PSF model for WFE retrieval that exploits the spatial diversity of the PSF in wide-field imagers using exclusively in-focus observations.
Studies such as \cite{dean2006phase} combine a parametric model with non-parametric optimisation to estimate the WFE for a single pointing using defocused measurements. 
% WaveDiff extends this semi-parametric approach to modelling the WFE across the entire FOV using only in-focus stellar observations.
WaveDiff takes a different approach to phase retrieval, using only in-focus observations and modelling the WFE across the entire FOV.
In the wide-field setting, WaveDiff \citep{liaudat2023} stands out as the only demonstrated semi-parametric method capable of modelling the chromatic variation of the PSF across the entire FOV solely from in-focus observations. 
We refer the reader to \citet{liaudat2023_2} for a review on PSF modelling for WL studies.

A growing body of work is exploiting differentiable optics and automatic differentiation \citep{margossian2019review}. 
This methodology is being developed for different purposes like direct exoplanet imaging \citep{feng2025exoplanet}, design of phase masks \citep{Wong2021}, design of diffractive pupils \citep{desdoigts2024}, phase retrieval \citep{desdoigts2023differentiable} and a calibration for the JWST \citep{gardner2006james} aperture masking interferometer system \citep{desdoigts2025amigo}. 
However, the methods developed are focused on alternative applications and cannot be easily reused for modelling the PSF in wide-field surveys, where modelling spatial and chromatic variations is essential. 

% \begin{table*}
% \centering
% \caption{Comparison of relevant PSF and phase retrieval methods across key features. 
% \cmark~indicates the feature is present, \xmark~indicates it is absent, 
% and \pmark~indicates partial support.}
% \label{tab:method_comparison}
% \begin{tabular}{lcccccc}
% \hline\hline
% Method & In-focus only & Full FOV & Semi-parametric & Chromatic PSF & No prior optical & Accurate WFE \\
%        &               &          &                 & modelling     & design required  & recovery \\
% \hline
% Gonsalves (1976)        & \cmark & \xmark & \xmark & \xmark & \cmark & \xmark \\
% Fienup (1993)           & \xmark & \xmark & \cmark & \xmark & \cmark & \cmark \\
% Dean et al. (2006)      & \xmark & \xmark & \cmark & \xmark & \cmark & \cmark \\
% Jurling \& Content (2012) & \cmark & \cmark & \xmark & \pmark & \xmark & \cmark \\
% WaveDiff (Liaudat et al. 2023a) & \cmark & \cmark & \cmark & \cmark & \cmark & \xmark \\
% This work               & \cmark & \cmark & \cmark & \cmark & \cmark & \cmark \\
% \hline
% \end{tabular}
% \end{table*}

Phase retrieval is an inverse problem that involves estimating the phase of a complex signal after a measurement process has lost phase information by measuring the amplitude of the signal. 
This problem is often encountered in physics, particularly in optics, where, for example, a CCD chip measures the intensity (pixel PSF), rather than the electric field phase (WFE) of the arriving signal. 
The phase retrieval problem gained traction in the scientific community, motivated by the severe aberrations of the Hubble Space Telescope (HST) mirror. 
Phase retrieval algorithms were developed to infer and correct HST's aberrations \citep{fienup1993}. 
Several algorithms have been developed to tackle this problem, including the Gerchberg-Saxton algorithm \citep{Gerchberg1972APA}, compressed sensing \citep{candes2013}, and deep learning \citep{metzler2018prdeep, icsil2019deep}. 
See \citet{dong2023phase} for a review on phase retrieval methods. The standard scenario for phase retrieval aims to estimate the phase of a complex signal from several intensity measurements of the same signal. 
In our WFE estimation problem, we are in a different situation where we have a single degraded intensity observation (pixel PSF) for each underlying complex electric field phase (WFE). 
However, we have multiple WFE representations that are linked to the same PSF field. 
This scenario makes the standard algorithms inapplicable for our problem.

%=======================================================================================
\section{The WaveDiff PSF model}
\label{sec:wavediff}
%=======================================================================================
WaveDiff is a semi-parametric framework that models the PSF in wavefront space. 
Its formulation relies on a WFE representation that is propagated through an optical differentiable forward model, enabling the generation of the corresponding PSF in the focal plane.
WaveDiff captures the spatial and spectral variation of the PSF and uses observations of unresolved stars, which are considered point sources, to characterise the instrumental response of a given telescope. 
Each stellar observation samples the PSF field at a different position in the FOV, and by using them jointly, WaveDiff can constrain the underlying PSF field. 
As a result, the quality of the reconstruction depends on the number of available stars.
During training\footnote{"Training" refers to the joint optimisation of both the model parameters and the non-parametric features, and should not be confused with deep learning network training.}, WaveDiff takes two inputs: the position of the observed stars $(u_i, v_i)$, expressed in FOV coordinates, and their corresponding SEDs, which are used for spectral integration as in the case of single-band optical telescopes. 
The predicted PSFs $\bar{H}(u_i, v_i)$ are compared with the observed stars images $\bar{I}(u_i, v_i)$ through a loss function, and fitting WaveDiff to the data consists of optimising its parameters $\theta$ to minimise this function (see \autoref{subsec:wf_optim}). 
Once optimised, WaveDiff is able to predict the PSF of the telescope at new positions in the FOV. 

The WaveDiff WFE model consists of two contributions: a parametric term, $\Phi^{\text{Z}}$, and a non-parametric (NP) term, $\Phi^{\text{NP}}$. 
The total WFE representation is defined as the sum of both terms as follows,
\small
\begin{multline}
    \Phi_{\theta}(\bar{x}, \bar{y} |u_i, v_i) = \underbrace{\sum_{l=1}^{n_{\text{Z}}} f^{\text{Z}}_l(u_i, v_i) \; S^{\text{Z}}_l(\bar{x},\bar{y})}_{\Phi^{\text{Z}}(\bar{x}, \bar{y} |u_i, v_i)} \;\;+ \\
    \underbrace{\sum_{k=1}^{n_{\text{NP}}} f^{\text{NP}}_k(u_i, v_i) \; S^{\text{NP}}_k(\bar{x},\bar{y})}_{\Phi^{\text{NP}}(\bar{x}, \bar{y} |u_i, v_i)} \;.
    \label{eq:wfe_psf_model}
\end{multline}
\normalsize
The WFE is expressed as a linear combination of wavefront features, $S^{\text{Z}}_l(\bar{x},\bar{y})$ and $S^{\text{NP}}_k(\bar{x},\bar{y})$ $\in \mathbb{R}^{K \times K}$, which are shared across the entire FOV. 
The spatial variation of the WFE is determined by the weights associated with each feature, computed through the spatial functions $f^{\text{Z}}_l(u_i,v_i)$ and $f^{\text{NP}}_k(u_i,v_i)$.
Further details on WFE model are provided in the following subsection.

In WaveDiff, the WFE is propagated through a forward differential model that emulates the optical system of the telescope and the degradations related to the detector.
The forward model is based on an approximation of the optical model of the telescope to a single converging lens \citep[\S 2.2]{liaudat2023_2}. 
In such a case, it is possible to relate the propagation and diffraction of the electric field between the pupil plane and the focal plane through the Fresnel diffraction approximation \citep{goodman2005}. 
Given a position in the FOV $(u_i, v_i)$, the wavelength-dependent electric field in the pupil plane is written as
\begin{equation}
    \mathcal{U}_{\text{p}}(x,y;\lambda|u_i,v_i) = \mathcal{P}(x,y) \exp{\left[\frac{2\pi j}{\lambda} \Phi_{\theta}(x,y|u_i, v_i) \right]},
\end{equation}
where $\mathcal{P}\;:\;\mathbb{R}^2 \rightarrow [0, 1]$ represents the obscuration encountered at the pupil plane due to occultation and superposition of mirrors and supports of the optical system of the telescope. 
The monochromatic PSF in the focal plane is calculated as follows,
\small
\begin{align}
    \mathcal{H}(u,v;\lambda|u_i, v_i) = \frac{a}{\lambda f_{\rm L}} \bigg|\bigg| &\iint_{-\infty}^{+\infty} \mathcal{U}_{\text{p}}(x,y;\lambda|u_i,v_i) \nonumber \\
    & \exp{\bigg[-\frac{2 \pi j}{\lambda f_{\rm L}} (ux+vy) \bigg]} \textrm{d}x \textrm{d}y \bigg|\bigg|^2,
    \label{eq:forward_model}
\end{align}
\normalsize
where $a$ is a constant amplitude that does not depend on the optical system and $f_{\rm L}$ is the focal length of the telescope.
Further details of the observation model can be found in \citet{liaudat2023}.

\subsection{Parametric and non-parametric parts}
%===============================================
Let us first recall how the parametric and non-parametric parts of the WaveDiff model are defined.
The parametric part is given by
\begin{equation}
    \Phi^{\text{Z}}(\bar{x},\bar{y};\lambda|u, v) =  \sum_{l=1}^{n_{\text{Z}}} \underbrace{\bm{\pi}^{\text{Z}}_{l}(u, v)^{\text{T}} \, \mathbbm{1}_{n_{d_{\text{Z}}}}}_{f_{l}^{\text{Z}}(u, v) } \, S_{l}^{\text{Z}}(\bar{x}, \bar{y}) \,,
\end{equation}
where $S_{l}^{\text{Z}}$ is the $l$-th Zernike polynomial (see \cite{noll1976} for the definition), $\mathbbm{1}_{n_{d_{\text{Z}}}} \in \mathbb{R}^{n_{d_{\text{Z}}} \times 1}$ is a vector of ones, $f_{l}^{\text{Z}}(u,v)$ is the polynomial of the index $l$, and for each $l$ we have a different polynomial $\bm{\pi}_{l}^{\text{Z}}$ that is given by
\begin{equation}
    \bm{\pi}^{\text{Z}}_{l}(u,v) = \left[\pi^{\text{Z}}_{l,[0,0]},\, \pi^{\text{Z}}_{l,[1,0]} u,\, \pi^{\text{Z}}_{l,[0,1]} v,\, \cdots \,,\, \pi^{\text{Z}}_{l,[0,d_{\text{Z}}]} v^{d_{\text{Z}}} \right] \,,
 \label{eq:polynomial_Zernike}
\end{equation}
where we are using a Zernike order of $n_{\text{Z}}$, and a maximum polynomial degree of $d_{\text{Z}}$ that gives us $n_{d_{\text{Z}}}$ monomials. 
The non-parametric part is given by
\begin{equation}
    \Phi^{\text{NP}}(\bar{x},\bar{y};\lambda|u,v) =  \sum_{k=1}^{n_{\text{NP}}} \underbrace{\bm{\pi}^{\text{NP}}(u, v)^{\text{T}} \, \bm{a}_{k}}_{f^{\text{NP}}_k(u, v)} \, S_{k}^{\text{NP}}(\bar{x}, \bar{y}) \, ,
 \label{eq:dd_wfe_model}
\end{equation}
where $S_{k}^{\text{NP}}$ are the non-parametric features, which are completely learned from observations, unlike Zernike polynomials, which are pre-calculated. $\bm{a}_{k}$ is a column of the mixing matrix $A$, and $\bm{\pi}^{\text{NP}}(u, v)$ is a FOV position polynomial shared by all the NP features. 
The polynomial is given by
\begin{align}
    \bm{\pi}^{\text{NP}}(u, v) = \Big[ \pi^{\text{NP}}_{[0,0]},\, &\pi^{\text{NP}}_{[1,0]} u ,\, \pi^{\text{NP}}_{[0,1]} v ,\, \pi^{\text{NP}}_{[2,0]} u^{2} ,\, \nonumber \\
	& \pi^{\text{NP}}_{[1,1]} u v ,\, \cdots ,\, \pi^{\text{NP}}_{[0, d_{\text{NP}}]} v^{d_{\text{NP}}} \Big]^{\text{T}} .
    \label{eq:polynomial_dd}
\end{align}
It is important to note that, if the mixing matrix $A$ is diagonal, then each NP feature contributes to a single monomial from $\bm{\pi}^{\text{NP}}$. 
However, as $A$ is also optimised, which makes the model more flexible, there is no reason for it to be diagonal. 
Therefore, the NP features will contribute to more than one monomial.

\subsection{Optimising the WaveDiff model}
\label{subsec:wf_optim}
%=========================================
Given the parameterisation of the PSF field in wavefront space presented in \autoref{eq:wfe_psf_model}, the optimisation of WaveDiff consists of finding the set of parameters $\theta$ that best fit the data.
The optimisable parameter set $\theta$ includes the following elements: the polynomial coefficients of the wavefront feature spatial distribution, both parametric ($\pi_l^\text{Z}[i,j]$ for $i+j \leq d_\text{Z}$ and $l=1,2,\ldots,n_\text{Z}$) and non-parametric ($\pi^\text{NP}[i,j]$ for $i+j \leq d_\text{NP}$); the NP feature mixing matrix $A\in\mathbb{R}^{n_\text{NP}\times n_\text{NP}}$ (where $n_\text{NP}=(d_\text{NP}+1)(d_\text{NP}+2)/2$); and the NP wavefront features themselves, $S^\text{NP}_k$.
In a data-driven\footnote{Data-driven refers to components whose features are learned directly from data rather than predefined analytically.} approach, we optimise the parameters that model the WFE based on observations.
% The key problem is that the ground truth (GT) WFE is not directly accessible; instead, the available data consist of observations in pixel space. 
The forward model that maps wavefront space to pixel space is given in \autoref{eq:forward_model}. 
To reproduce the observation of a star, it is further necessary to spectrally integrate the PSF with the corresponding stellar SED. %, as described in \autoref{eq:obs_star}. 
We define the optical forward model that predicts the observation of a star at a given $(u,v)$ position in the FOV, combining the two equations introduced above
\begin{equation}
    \label{eq:FWD}
    \widehat{\bar{H}}(u_i,v_i)=\mathtt{Fwd} \bigg\{\text{SED}_{(u_i,v_i)}, \; \Phi_\theta(u_i,v_i) \bigg\}.
\end{equation}
Given that the available data consist of stellar observations in pixel space, the optimisation problem can be formulated in pixel space as
\begin{equation}
    \hat{\theta}=\arg\min_{\theta \in \Theta} \frac{1}{n_\text{stars}} \sum_{i=1}^{n_\text{stars}} \frac{1}{\hat{\sigma}_i}\left|\left|\widehat{\bar{H}}(u_i,v_i) - \bar{I}(u_i,v_i) \right|\right|^{2}_{F},
    \label{eq:optim_pixel}
\end{equation}
where $\hat{\bar{H}}$ denotes the predicted star observation, $\bar{I}$ the corresponding measured observation, and $\hat{\sigma}_i$ is the estimated observational noise of each star.

To solve this problem, WaveDiff employs a stochastic gradient descent algorithm \citep{kingma2014adam}. 
This approach is feasible because the forward model is implemented in a differentiable framework like TensorFlow \citep{tensorflow2015} such that its output is differentiable with respect to the parameters of the model. 
Automatic differentiation greatly facilitates this process. 
For optimisation, WaveDiff adopts Rectified Adam \citep{liu2020}, a stochastic gradient-based method widely used in the machine learning community.

Each of the WaveDiff wavefront contributions, namely the parametric and the non-parametric components, are optimised in an alternating fashion while keeping the other fixed. 
Each optimisation cycle consists of a number of parametric epochs, governed by a parametric learning rate, followed by a number of non-parametric epochs, governed by a separate learning rate. 
The full optimisation process is composed of multiple cycles, where both the number of epochs and the learning rates can vary across cycles. 
Further details on the WaveDiff optimisation procedure and parameter initialisation are provided in \citet{liaudat2023}.

%=======================================================================================
\section{Wavefront recovery through model-based automatic differentiation}
\label{sec:methodology}
%=======================================================================================
Optimising the PSF model consists of finding the set of parameters $\theta$ that minimise the loss function given a set of observations. 
The underlying objective is to represent the WFE of the telescope at different positions in the FOV. 
However, we do not have any measurements of the WFE, but rather we observe its counterpart in pixel space. 
The observed PSF carries information about the corresponding WFE, yet since we measure the intensity of the light reaching the detector, all phase information is lost. 
Furthermore, as the stars are observed in-focus, the PSF collapses into a small region in the focal plane making the observation less informative to different WFE aberrations. 
This is the motivation for using out-of-focus images \citep{davis2016, Roodman2014}, usually referred to as doughnuts, in WFE estimation as they are more spread out in the focal plane and more informative. 
For these reasons, phase recovery from in-focus observations is extremely difficult. 

\begin{figure}
    \centering
    \includegraphics[width=\linewidth,trim={.5cm 3.8cm .5cm 0},clip]{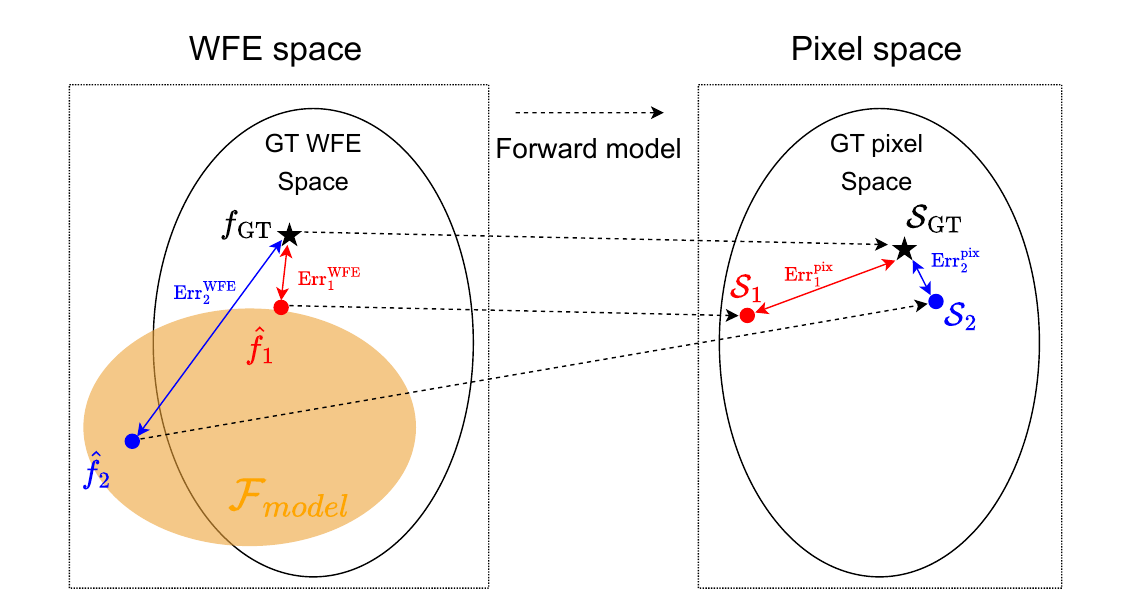}
    \caption{\small Schematic of the WaveDiff optimisation scenario in \protect\citet{liaudat2023}.
    The black ovals denote the sets of physically feasible solutions in WFE space and pixel space. The ground-truth solution $f_{\text{GT}}$ in WFE space, and $\mathcal{S}_{\text{GT}}$ in pixel space, is indicated by a black star. 
    The orange region represents the set of WaveDiff solutions, $\mathcal{F}_{\text{model}}$, which does not include $f_{\text{GT}}$. 
    The solution $\hat{f}_1 \in \mathcal{F}_{\text{model}}$ minimises error in wavefront space, and its counterpart $\mathcal{S}_{1}$ is shown in the pixel space.
    Minimising the error in pixel space yields solution $\mathcal{S}_{2}$, which has a lower pixel-space error than $\mathcal{S}_{1}$; however, its counterpart $\hat{f}_{2}$ exhibits a much higher error in wavefront space.
    }
    \label{fig:scenario_1}
\end{figure}

WaveDiff is optimised by minimising the loss in pixel space, as described in \autoref{eq:optim_pixel}, and given the degeneracies of the pixel space with respect to the wavefront space there is no guarantee that the solution to this optimisation problem will also retrieve the correct WFE. 
The WaveDiff optimisation scenario considered in \citet{liaudat2023} is illustrated in \autoref{fig:scenario_1}.
In this scenario, we assume that WaveDiff is not complex enough to represent the underlying PSF model in the wavefront space.
On the left side we present the space of all possible solutions in the wavefront space, and on the right side in the pixel space. 
These two spaces are related through the forward model. 
The black oval represents the set of real, physically possible solutions, among which is the ground-truth target function $f_{\text{GT}}$. 
Its pixel space counterpart is denoted by $\mathcal{S}_{\text{GT}}$. 
The family of solutions represented by WaveDiff, denoted $\mathcal{F}_{\text{model}}$, is contained in the orange region. 
We note that WaveDiff is capable of representing solutions that are not physically feasible in wavefront space. 
This is due to the high number of degrees of freedom provided by the learnable non-parametric features $S_{k}^{\text{NP}}$. 
In this diagram, $\hat{f}_1$ represents the closest WFE model to the ground truth WFE $f_{\text{GT}}$, within the family of solutions provided by WaveDiff. 
However, its counterpart $\mathcal{S}_{1}$ is not necessarily the solution to the optimisation problem in pixel space. 
In this scenario, the solution we retrieve in pixel space is $\mathcal{S}_{2}$, which has a lower error in such a space compared to $\mathcal{S}_{1}$. 
Yet, this solution has a higher error in wavefront space and may even be a non-physically plausible solution.
We also note that under the aforementioned hypothesis we cannot identify, if exists, the global minimum in the WFE space. 
Given the non-convexity of the optimisation problem and the limitations of the optimisation algorithm, we converge to a local minimum.

We propose, in this section, a novel procedure to optimise the WaveDiff PSF model. 
The section is based on a scenario where the parametric part of the model can reproduce the GT WFE field. 
Under this hypothesis, we know at least one global minimum solution, which is the GT WFE. 
However, estimating the WFE parameters from degraded in-focus observations is not possible with the current WaveDiff optimisation procedure. 
The proposed method introduces wavefront feature projections relating WaveDiff's parametric and non-parametric parts, better exploiting the physical knowledge of the PSF field and the flexibility of the non-parametric model. 

\subsection{Hypothesis and motivations}
%======================================
As discussed in \autoref{sec:methodology} the current WaveDiff optimisation procedure converges to a local minimum. 
This observation is expected as we tackle a non-convex problem with a gradient-based optimisation method. 
The optimisation consists in cycles where we optimise the parametric part and then continue with the non-parametric part. 
See \autoref{al:wavediff_training_original} in the \autoref{apx:algorithms} for more details. 
We have used two optimisation cycles in all the previous experiment with the WaveDiff model. 
We noted that the model was stuck in a local minimum that the optimisation could not escape. 
This fact is easily confirmed by plotting the validation loss, which is the loss computed on the testing datasets, as a function of the number of epochs. 
Therefore, even if we increase the number of cycles in the optimisation procedure, the algorithm will not escape from the local minimum as we will later show.

\begin{figure}
    \centering
    \includegraphics[width=\linewidth,trim={1cm 2.7cm 1.5cm 0},clip]{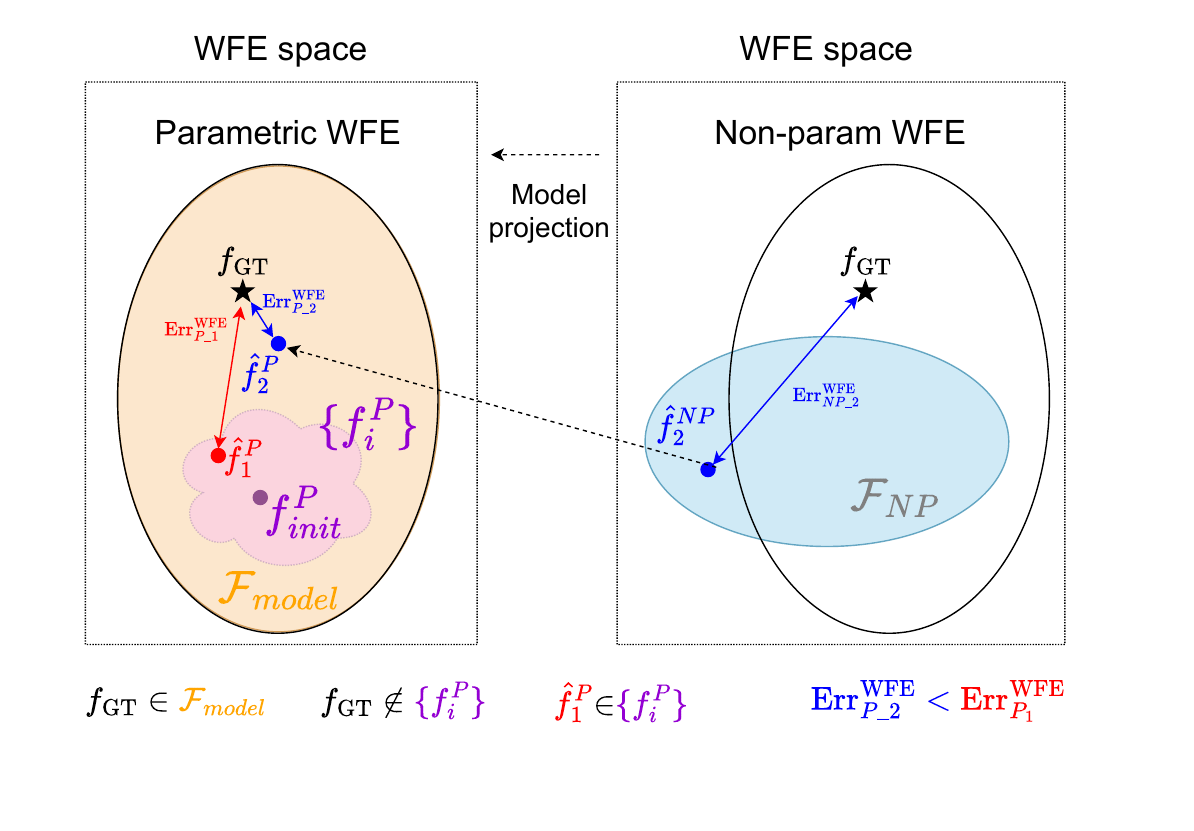}
    \caption{\small Schematic of the proposed WaveDiff optimisation scenario.
    The left side shows the set of parametric solutions $\mathcal{F}_{\text{model}}$, in orange, which includes the ground-truth solution $f_{\text{GT}}$, indicated by a black star. 
    Starting from a solution $f_{\text{init}}^\text{P}$, the set of possible outcomes obtained through parametric-only optimisation  $\{f_i^\text{P}\}$ is shown in purple. 
    One parametric-only solution, $\hat{f}_1^\text{P}$, is shown in red.
    The set of non-parametric solutions, $\mathcal{F}_{\text{NP}}$, is displayed on the right-hand side in light blue. 
    The non-parametric solution $\hat{f}_2^{\text{NP}}$ is projected onto the parametric model, yielding $\hat{f}_2^{\text{P}}$, whose error in WFE space is smaller than that of $\hat{f}_1^{\text{P}}$.
    }
    \label{fig:scenario_2}
\end{figure}

In the original WaveDiff scenario, the authors assumed the parametric model could not fully represent the GT WFE field. 
They showed that the non-parametric part could estimate a useful representation of the WFE field, which had low errors in the pixel space even if it had high errors in the WFE space (see \autoref{fig:scenario_1}). 
The alternative scenario to consider is that parametric part is complex enough to reproduce the GT WFE field. 
This scenario is illustrated in \autoref{fig:scenario_2}.
The family of parametric models $\mathcal{F}_{\text{model}}$ is represented on the left side by the orange set, which includes the ground-truth model $f_{\text{GT}}$. 
Due to the complexity of the parametric model optimisation problem, starting from an initial random solution $f_{\text{init}}^{\text{P}}$ typically leads only to a restricted subset of solutions, $\{f_i^{\text{P}}\}$, which generally exhibit poor performance in terms of pixel and WFE metrics.
Even when the initialisation is close to the ground truth, the stochastic optimisation process of the parametric model often diverges from this favourable starting point.
Consequently, both the WFE error, $\text{Err}_{\text{P}\_1}^{\text{WFE}}$, and the pixel error $\text{Err}_{\text{P}\_1}^{\text{pix}}$ of the parametric-only solution $\hat{f}_1^{\text{P}}$ increase. 
This behaviour highlights the inherent difficulty of optimising the parametric component using in-focus, degraded star observations.

On the right-hand side of \autoref{fig:scenario_2}, the set of non-parametric solutions is shown in light blue. 
By itself, the non-parametric model cannot fully represent the ground-truth model, but it can reach solutions beyond the scope of the parametric model.
The non-parametric component has been effective in converging to a good pixel-level representation of the PSF field, even when the WFE error remains high, and its optimisation is considerably more stable than that of the parametric counterpart.
We attribute this stability to the over-parametrisation of the non-parametric component, which results in a much smoother—although still non-convex—optimisation landscape \citep{choromanska2015}.
For context, in the experiment of \citet{liaudat2023}, the parametric part of the \textit{WaveDiff} model had $90$ parameters, whereas the non-parametric part had approximately $1.3 \times 10^{6}$.

Taking all the aforementioned points into account, we aim to exploit the more stable optimisation of the non-parametric component together with the ability of the parametric component to represent the ground-truth WFE field.
We propose to optimise the non-parametric part of the model and then transfer the information it shares with the parametric part.
This approach is illustrated in blue in \autoref{fig:scenario_2}, where the non-parametric solution $\hat{f}_2^{\text{NP}}$ is projected onto the parametric model, yielding the solution $\hat{f}_2^{\text{P}}$ whose WFE error ($\text{Err}_{\text{P}\_2}^{\text{WFE}}$ significantly smaller than that of the previous parametric-only solution $\hat{f}_1^{\text{P}}$ (shown in red).
In the following, we describe in detail how the non-parametric component is transferred, or projected, onto the parametric one.

\subsection{Wavefront features projection}
%=========================================
To transfer information from the non-parametric to the parametric part, we need an operator that links non-parametric wavefront features and Zernike polynomials. 
We propose a projection operator that decomposes an arbitrary WFE map into Zernike contributions up to the $n$th order, together with a higher-order WFE residual. 
This decomposition enables us to identify the non-parametric contribution that will be injected into the parametric part.
However, the construction of this operator is not straightforward as the presence of the obscuration mask breaks the orthogonality of the Zernike basis, preventing the direct use of the naïve inner product to define the WFE projection, as this would bias the resulting decomposition. 
We therefore propose a method of iterative projections that compensates for the lack of orthogonality of obscured Zernike features.

\subsubsection{Zernike projection}
%=================================
We define the naïve projection between the wavefront features $S_k, S_l \in \mathbb{R}^{K \times K}$ as follows
\begin{equation}
    \langle S_k(\bar{x},\bar{y}) \,,\, S_l(\bar{x},\bar{y}) \rangle = \sum_{\bar{x},\bar{y} \in D} \frac{S_{k}(\bar{x},\bar{y}) \, S_{l}(\bar{x},\bar{y})}{\left| D \right|} ,
    \label{eq:inner_product}
\end{equation}
where $D \in \mathbb{R}^{K \times K}$ is a circular aperture, and $|D|$ is the amount of non-zero elements in $D$. 
Since Zernike polynomials are orthogonal on the unit disk by construction, this definition guarantees the orthonormality between the Zernike features, which reads
\begin{equation}
    \langle S_{k}^{\text{Z}}(\bar{x},\bar{y}) \,,\, S_{l}^{\text{Z}}(\bar{x},\bar{y}) \rangle = \delta_{k \,l} \,,
\end{equation}
and allows us to define a projection operator over the Zernike basis
\begin{equation}
    \text{P}_{S^Z_k}\big(\Phi(\bar{x},\bar{y})\big) = \big\langle \Phi(\bar{x},\bar{y}), S^Z_k(\bar{x},\bar{y}) \big\rangle .
    \label{eq:z_proj}
\end{equation}
Consequently, if a WFE map is a linear combination of Zernike features, which is written as
\begin{equation}
    \text{WFE}(\bar{x},\bar{y}|u,v) = \sum_{k=1}^{n_{\text{Z}}} f^{k}(u,v) \, S_{k}^{\text{Z}}(\bar{x},\bar{y}),
\end{equation}
the Zernike coefficients can be retrieved by means of the projection operator as follows
\begin{equation}
    f^{k}(u,v) = \text{P}_{S_k^Z}\big(\text{WFE}(\bar{x},\bar{y}|u,v) \big) \,.
\end{equation}

\subsubsection{Non-parametric WFE projection}
\label{sec:non-param_projection}
%============================================
The WFE representation has two components of very different nature: a parametric contribution built from well defined Zernike polynomials and a non-parametric contribution whose features are learned throughout the optimisation procedure. 
Given the data-driven nature of the second features, we expect that they will not be orthogonal to the Zernike basis, resulting in an overlap between the parametric and non-parametric contributions. 
We can thus decompose each non-parametric feature into two orthogonal components
\begin{equation}
    S^{\text{NP}}_i(\bar{x}, \bar{y}) = S_{<N_\text{Z}}^{^Z}(\bar{x}, \bar{y}) +  S^{\mathcal{O}}(\bar{x}, \bar{y}),
\end{equation}
where the first component is a weighted sum of Zernike features up to order $N_\text{Z}$ and the second component is a higher order residual. 
The objective of the wavefront projection algorithm is to allow the first term to be represented by the parametric contribution, leaving only the higher orders to the non-parametric model.

We use the projection operator defined in \autoref{eq:z_proj} to decompose the NP part with respect to a Zernike feature of index $l$ as follows
\begin{equation}
    \text{P}_{S^Z_l}\big( \Phi^{\text{NP}} \big) = \sum_{k=1}^{n_{\text{NP}}} \bm{\pi}^{\text{NP}}(u,v)^{\text{T}} \; \bm{a}_{k} \; \langle S_{k}^{\text{NP}}, \; S_{l}^{\text{Z}} \rangle.
    \label{eq:WFE_Zernike_inner_product}
\end{equation}
If we perform the same projection with the parametric part, we obtain 
\begin{equation}
    \text{P}_{S^Z_l}\big( \Phi^{\text{Z}} \big) = \bm{\pi}^{\text{Z}}_{l}(u,v)^{\text{T}} \, \mathbbm{1}_{n_{d_{\text{Z}}}} = f_{l}^{\text{Z}}(u,v) \,.
\end{equation}
We reformulate \autoref{eq:WFE_Zernike_inner_product} as follows
\begin{equation}
    \text{P}_{S^Z_l}\big( \Phi^{\text{NP}} \big) = \underbrace{\sum_{\substack{i,j \geq 0 \\ i+j \leq d_{\text{Z}}}} \Delta \pi_{l,[i,j]}^{\text{Z}} \, u^{i} \, v^{j} }_{\Delta f_{l}^{\text{Z}}(u,v)} + \underbrace{\sum_{\substack{i+j > d_{\text{Z}} \\ i+j \leq d_{\text{NP}}}} c_{[i,j]} \, u^{i} \, v^{j} }_{C(u,v)} ,
\end{equation}
where $C(u,v)$ is a polynomial containing monomials of order greater than $d_{\text{Z}}$, and $\Delta f_{l}^{\text{Z}}(u,v)$ is a polynomial of degree $d_{\text{Z}}$ with parameters $\Delta \pi_{l,[i,j]}^{\text{Z}}$. 
We are assuming that $d_{\text{Z}} < d_{\text{NP}}$ which has been the case until now. Nevertheless, the results hold otherwise, because if $d_{\text{Z}} > d_{\text{NP}}$, then $C(u,v)=0$. 

We aim to find an expression for the coefficients $\Delta \pi_{l,[i,j]}^{\text{Z}}$, as they correspond to the part of the NP contribution that the parametric part can represent. 
These coefficients correspond to the parametric correction required by the parametric part to acquire the information from the NP part. 
Rearranging \autoref{eq:WFE_Zernike_inner_product} using linear algebra we obtain
\begin{align}
    \label{eq:WFE_Zernike_inner_product_matrix}
    \text{P}_{S^Z_l}\big( \Phi^{\text{NP}} \big) = &
    \begin{bmatrix}
        \pi^{\text{NP}}_{[0,0]} & \cdots & \pi^{\text{NP}}_{[0,d_{\text{NP}}]} \, y^{d_{\text{NP}}}
    \end{bmatrix} 
	\\
    & \left(
    \begin{bmatrix}
        \vert \\
        \bm{a}_{1} \\ 
        \vert \\
    \end{bmatrix}
    \underbrace{\left\langle S_{1}^{\text{NP}}, \; S_{l}^{\text{Z}} \right\rangle}_{\in \mathbb{R}}
    + \cdots +
    \begin{bmatrix}
        \vert \\
        \bm{a}_{n_{\text{NP}}} \\ 
        \vert \\
    \end{bmatrix}
    \underbrace{\left\langle S_{n_{\text{NP}}}^{\text{NP}}, \; S_{l}^{\text{Z}} \right\rangle}_{\in \mathbb{R}}
    \right) .\nonumber
\end{align}
Then, the non-parametric corrections for the first two terms, $\Delta \pi_{l,[0,0]}^{\text{Z}}$ and $\Delta \pi_{l,[1,0]}^{\text{Z}}$, can be expressed as
\begin{align}
    \Delta \pi_{l,[0,0]}^{\text{Z}} =& \pi_{[0,0]}^{\text{NP}} \Big( \bm{a}_{1}[1] \; \langle S_{1}^{\text{NP}}, \, S_{l}^{\text{Z}} \rangle \nonumber \\
	&+ \bm{a}_{2}[1] \; \langle S_{2}^{\text{NP}}, \, S_{l}^{\text{Z}} \rangle + \cdots + \bm{a}_{n_{\text{NP}}}[1] \; \langle S_{n_{\text{NP}}}^{\text{NP}}, \, S_{l}^{\text{Z}} \rangle \Big) \in \mathbb{R}, \nonumber \\
    \Delta \pi_{l,[1,0]}^{\text{Z}} =& \pi_{[1,0]}^{\text{NP}} \Big( \bm{a}_{1}[2] \; \langle S_{1}^{\text{NP}}, \; S_{l}^{\text{Z}} \rangle \nonumber \\
	&+ \bm{a}_{2}[2] \; \langle S_{2}^{\text{NP}}, \, S_{l}^{\text{Z}} \rangle + \cdots + \bm{a}_{n_{\text{NP}}}[2] \; \langle S_{n_{\text{NP}}}^{\text{NP}}, \, S_{l}^{\text{Z}} \rangle \Big) \in \mathbb{R}, \nonumber
\end{align}
where $\bm{a}_{1}[1]$ corresponds to the first element of the first column of the matrix $A$. 
By looking carefully to \autoref{eq:WFE_Zernike_inner_product_matrix}, we can generalise the parametric correction coefficients as follows
\begin{equation}
    \Delta \pi_{l,[i,j]}^{\text{Z}} = \pi_{[i,j]}^{\text{NP}} \sum_{n=1}^{n_{\text{NP}}} \bm{a}_{n}[p(i,j)] \; \text{P}_{S^Z_l}\big( S_{n}^{\text{NP}} \big),
    \label{eq:param_correction_terms}
\end{equation}
where $i,j \geq 0$ and $i + j \leq d_{\text{Z}}$, and $l = 1, \ldots, n_{\text{Z}}$. 
The order of the indexes $[i,j]$ follows the sequence presented in \autoref{eq:polynomial_dd}. 
This order is required to index the vectors $\bm{a}_n$, and is given by the following function
\begin{equation}
    p(i,j) = \frac{(i+j)(i+j+1)}{2} + j + 1.
\end{equation}

The required correction, ${\pi^{\text{Z}}}^{*}_{l,[i,j]}$, to apply to the parametric model in order to incorporate the information from the non-parametric part can be expressed as
\begin{equation}
\label{eq:parametric_update}
    {\pi^{\text{Z}}}^{*}_{l,[i,j]} = \pi_{l,[i,j]}^{\text{Z}} + \Delta \pi_{l,[i,j]}^{\text{Z}} \,.
\end{equation}
In \autoref{apx:projection}, we show how we can remove the projected part from the non-parametric model, thus ensuring that the total WFE remains unchanged.

\begin{algorithm*}
    \small
    \DontPrintSemicolon
    \caption{Zernike projection algorithm for obscured pupils}
    \label{al:projection_alg}
    \SetInd{0.2em}{2em} % Control indentation block width
    \textbf{Input:} $\Phi_{\theta}$, $N_{\text{max}}$, $\mathcal{P}$, $z_{\text{norm}}$, $N_{Z}$, $\{S_{l}^{Z}\}_{l=1}^{N_{Z}}$, $c, r = [0, \ldots, 0] \in \mathbb{R}^{N_{Z}}$ \\
    \textbf{Output:} $r$ \\ % \vspace{0.05in} \\ 
    \For{$n=0, 1, \ldots, N_{\text{max}}$}{
        \For{$i=0, 1, \ldots, N_{Z}$}{
            $c[\,i\,] \gets \text{P}_{S^Z_i}\big( \Phi_{\theta} \big) $ \tcp*{naïve projection on Zernike map}
            
        }
        \For{$j=0, 1, \ldots, N_{Z}$}{
            $\Phi_{\theta} \gets \Phi_{\theta} - \mathcal{P} \odot (c[j]*S_{j}^{Z}) $ \tcp*{remove the Zernike contribution with obscuration}
            $r[j] \gets r[j] + c[j]$ \tcp*{store contribution}
            $c[j] \gets 0$ \tcp*{set to zero Zernike coefficient before continuing}
        }
    } \vspace{0.05in}
    \Return $r$
\end{algorithm*}

\subsubsection{Iterative Zernike projection}
\label{sec:iterative_projection}
\begin{figure}
    \centering
    \includegraphics[width=0.48\textwidth,trim={0cm .4cm 0cm .cm},clip]{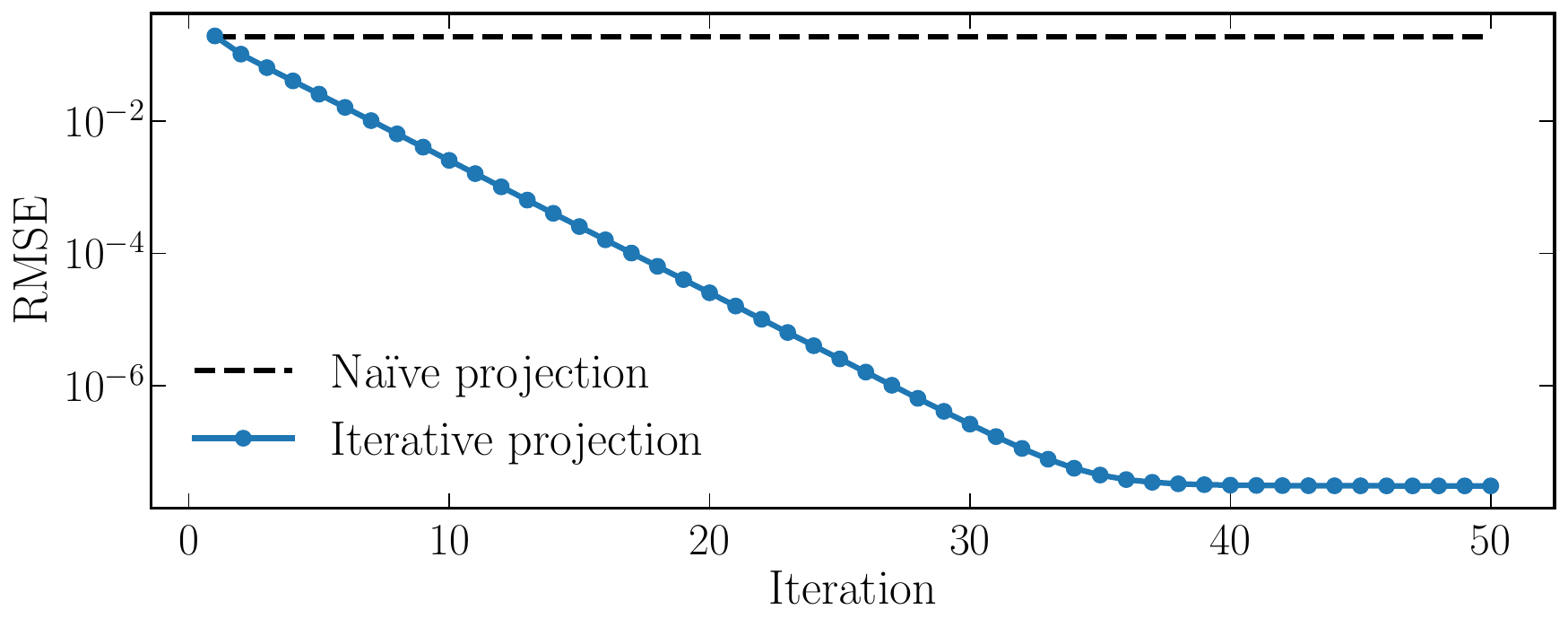}
    \caption{\small WFE reconstruction relative error for the naive and iterative projection algorithms.}
    \label{fi:zk_projection_comparison}
\end{figure}

\begin{figure}
    \centering
    \includegraphics[width=0.48\textwidth,trim={0cm .4cm 0cm 0cm},clip]{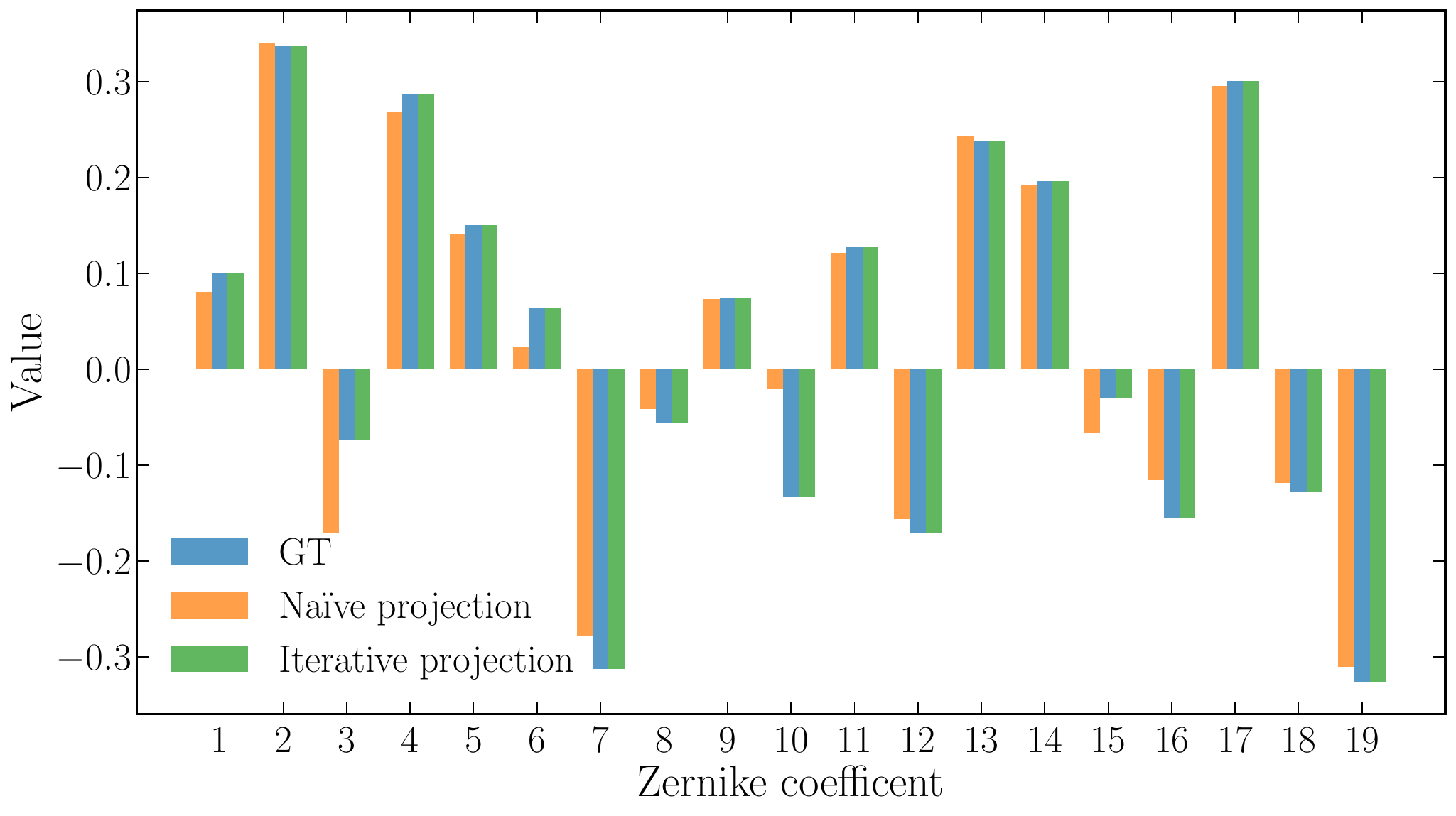}
    \caption{\small Estimated Zernike coefficients for the two projection algorithms, compared to the fiducial values (GT).}
    \label{fi:zk_projection_bar_plot}
\end{figure}

The presence of the obscuration mask in the WFE breaks the orthogonality of the Zernike basis
\begin{equation}
    \text{P}_{S_{l}^{\text{Z}}}\big( \mathcal{P}\odot S_{k}^{\text{Z}} \big) \neq \delta_{kl},
\end{equation}
thereby hindering the decomposition of the WFE onto Zernike polynomials. 
A solution to this problem could be to orthogonalise the Zernike polynomials over the obscured pupil. 
However, this solution is not feasible because, as the obscuration is caused by the shadow cast by mechanical supports of the telescope, it changes with the angle of incidence of the incoming light. 
That is, the obscuration varies across the FOV. Although WaveDiff does not yet model this spatial variation, we believe that such a solution would not be compatible with more realistic settings.
To address this, we adapt the projection algorithm to mitigate the Zernike mixing introduced by the obscuration mask. 
The proposed solution, presented in \autoref{al:projection_alg}, consists of successive naive WFE projections that gradually correct for non-orthogonality errors, allowing for the inclusion of a spatially varying obscuration mask.
To validate the projection algorithm, we constructed a fiducial WFE using Zernike polynomials up to order $n_\text{Z}=20$, including the obscuration mask from \cite{liaudat2023}. 
The WFE was then decomposed with the projection algorithm to obtain the estimated Zernike coefficients. From these coefficients, we reconstructed the WFE and compared it to the fiducial case.
\Autoref{fi:zk_projection_comparison} presents the relative error in WFE reconstruction for the naïve projection algorithm and for the new iterative algorithm over $50$ iterations. 
A substantial reduction in reconstruction error is achieved with the new algorithm, decreasing from approximately $10\%$ to below $10^{-7}$\%.
\Autoref{fi:zk_projection_bar_plot} displays the estimated Zernike coefficients from both algorithms against the fiducial values, further demonstrating the improved performance of the new projection method.

\subsection{Projection-based WFE recovery}
\label{sec:new_wfe_optim_algorithm}
%=========================================
Once the projection algorithm has been defined, we integrate this tool into the WaveDiff optimisation procedure with the objective of recovering the WFE and its spatial variation. 
It is important to emphasise that projecting the wavefront features does not alter the overall WFE representation; rather, it redistributes it between its two components, the parametric contribution $\Phi^{\text{Z}}$ and the non-parametric contribution $\Phi^{\text{NP}}$. 
Since each component is optimised separately, the projection repositions the optimisation points for the two contributions while preserving the total WFE. 
The modifications introduced in the WaveDiff optimisation process are detailed below:

\begin{figure}
    \centering
    \includegraphics[width=.9\linewidth,trim={0cm .2cm 0cm 0cm},clip]{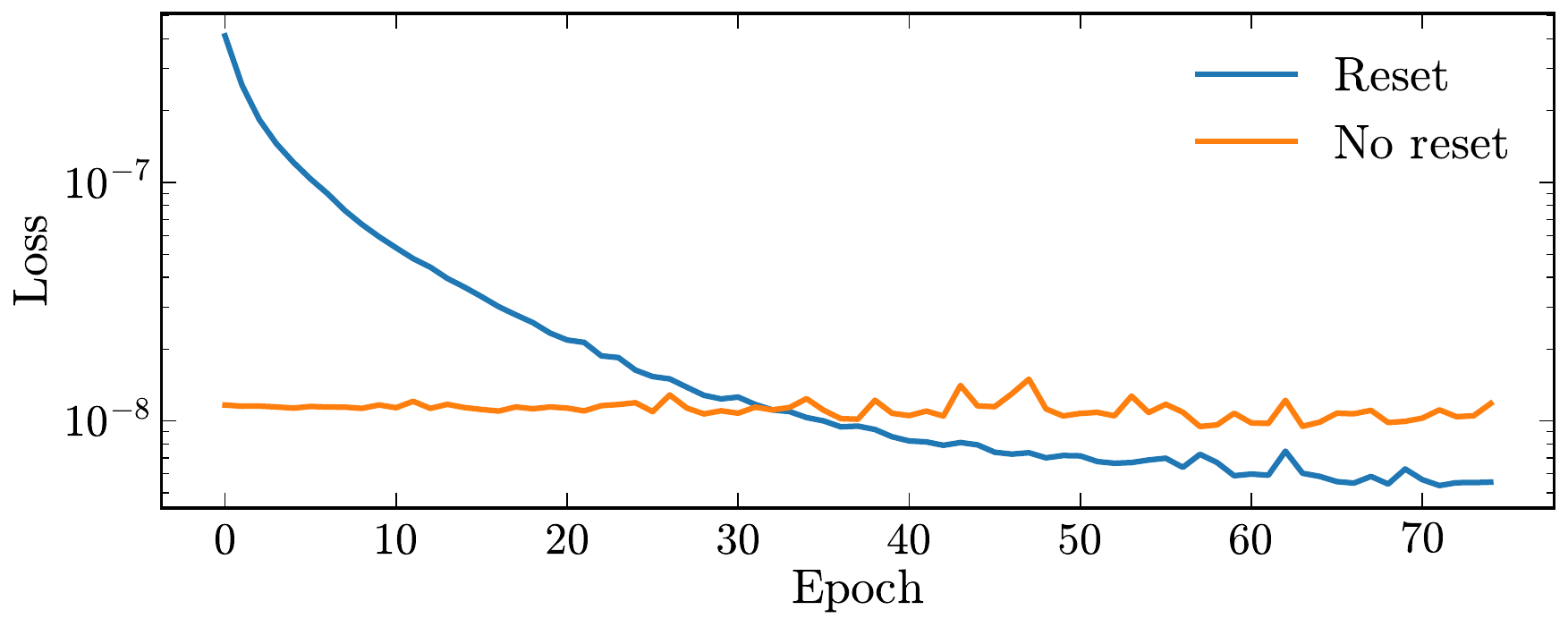}
    \caption{\small Loss function as a function of the number of epochs for the non-parametric part of the model in the second optimisation cycle. One model has been reset after performing the projection, and the other model has not been reset.}
    \label{fi:optim_curves_reset}
\end{figure}

\begin{itemize}
    \item[(i)] \textit{Wavefront features projection:} At the end of each optimisation cycle, which includes updating both the parametric and non-parametric components, wavefront features are projected from the non-parametric model onto the parametric model. This step relocates the optimisation points of the two contributions, with the main objective of guiding the parametric model closer to the global minimum. 
    \item[(ii)] \textit{Increased number of optimisation cycles:} Each new optimisation cycle starts from these relocated, and thus more favourable, optimisation points. To exploit this strategy, we increase the number of optimisation cycles from $2$ to $12$ cycles.
    \item[(iii)] \textit{Reinitialising the non-parametric part after projection:} Since the global WFE representation remains unchanged after projection, the optimisation process stalls if the model has converged to a local minimum. To address this issue, we propose reinitialising the non-parametric model after each projection. This step enables the model to escape a local minima and allows the subsequent cycle to explore new regions of the non-parametric optimisation landscape. \Autoref{fi:optim_curves_reset} shows the optimisation loss of the non-parametric part during the second cycle. The orange curve corresponds to a model that continues from the optimisation point of the previous cycle, whereas the blue curve corresponds to a model reinitialised at the end of the first cycle. In the former case, the model is stuck in a local minimum with no further improvement. In contrast, reinitialising the model restarts the optimisation process and yields a lower loss than in the non-reinitialised case.
    \item[(iv)] \textit{Ensuring optimisation convergence:} Increasing the number of optimisation cycles initially motivated us to reduce the number of epochs per cycle in order to limit the computational resources required for PSF model fitting. However, this reduction prevented the non-parametric model from reaching convergence, thereby reducing the effectiveness of the projection step. We therefore decided to maintain the original number of epochs per cycle, ensuring that the non-parametric model converges at the end of each cycle. The trade-off is an increase in computational cost.
    \item[(v)] \textit{Non-parametric-only optimisation:} The optimisation of the parametric component from degraded in-focus observations is challenging and highly degenerate. Even though the parametric model can, in principle, reproduce the ground-truth WFE field, estimating its parameters is not feasible with the gradient-based optimisation technique employed. We therefore focus on optimising the non-parametric component, which exhibits more stable behaviour due to its over-parameterisation. Until a more effective optimisation strategy for the parametric component is developed, we restrict its role to storing the projections of the non-parametric part at the end of each optimisation cycle.
\end{itemize}

The five points discussed above are incorporated into the optimisation procedure of the WaveDiff model, as detailed in \autoref{al:wavediff_training} in the Appendix. 
The following section presents the experiments conducted and the corresponding results.

%=======================================================================================
\section{Experimental results}
\label{sec:results}
%=======================================================================================
We test the proposed optimisation framework incorporating the wavefront feature projection. 
A fiducial PSF field is generated using the same simulator as in \citet{liaudat2023}. 
Following the approach outlined in point (v) of the previous section, we optimise only the non-parametric model and project it onto the parametric model at the end of each cycle. 
We analyse the effect of the number of cycles and the number of epochs per cycle on the optimisation process, as discussed in points (ii) and (iv). 
The optimised PSF model is evaluated on the test dataset, assessing the reconstruction both in pixel space and, since these are simulations, also in wavefront space. 
For each configuration, we run three simulations and report the median, the best run, and the standard deviation of the results.
In the following subsections, we describe the PSF field and star observation simulations, as well as the experiments conducted and their corresponding results.

\subsection{Simulations}
\label{sec:sims}
%=======================
\begin{figure}
    \captionsetup[subfigure]{justification=centering}
    \centering
    \begin{subfigure}[t]{.45\textwidth}
        \centering
        \includegraphics[width=\textwidth,trim={0cm 0 0cm 0cm},clip]{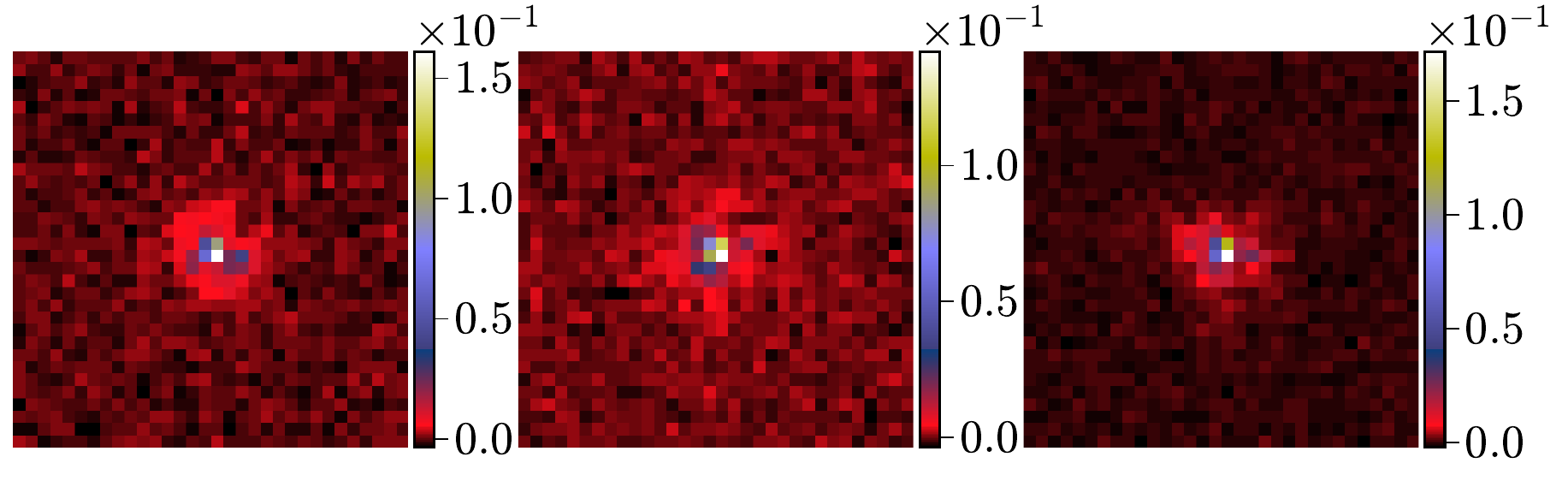}
        % \caption{}
        \label{fi:observation_example_a}
    \end{subfigure}\\
    \vspace{-.3cm}
    \begin{subfigure}[t]{.45\textwidth}
        \centering
        \includegraphics[width=\textwidth,trim={0cm 0 0cm 0cm},clip]{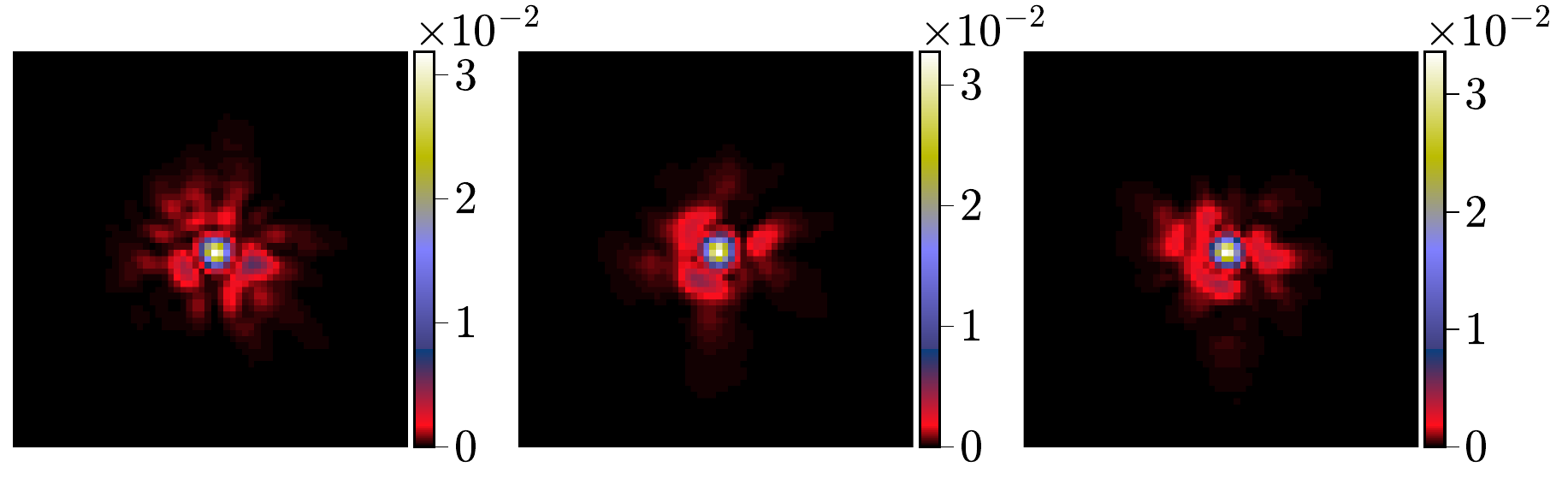}
        % \caption{}
        \label{fi:observation_example_b}
    \end{subfigure}\\
    \vspace{-.3cm}
    \begin{subfigure}[t]{.45\textwidth}
        \centering
        \includegraphics[width=\textwidth,trim={0cm .3cm 0cm 0cm},clip]{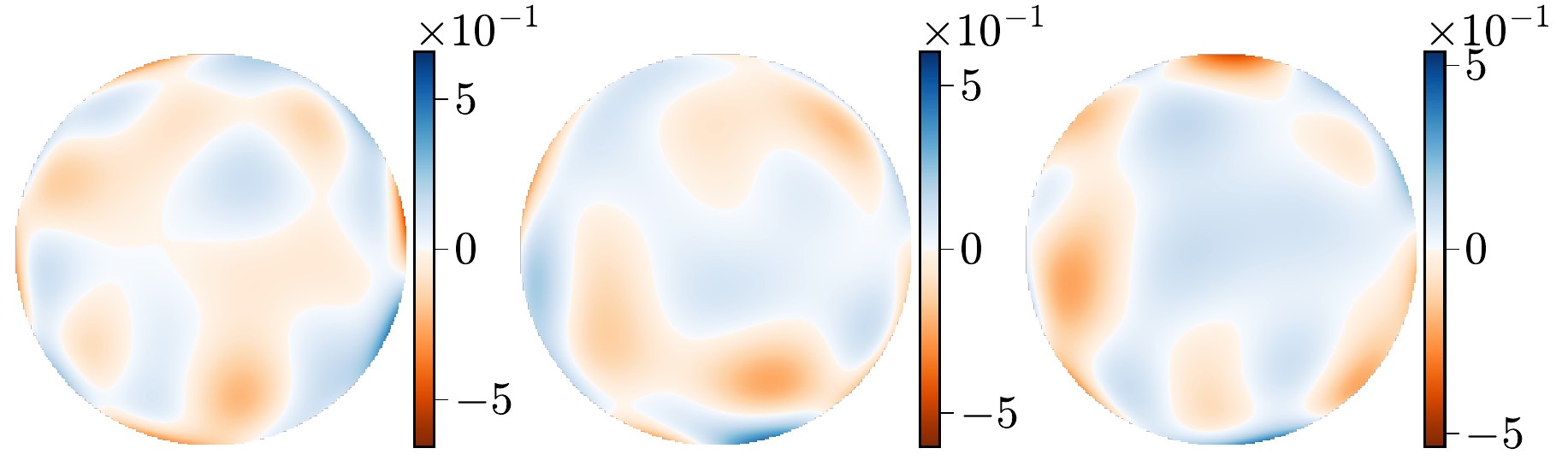}
        % \caption{}
        \vspace{-.5cm}
        \label{fi:observation_example_c}
    \end{subfigure}
    \caption{\small WaveDiff simulations at three different positions in the FOV. First row: the noisy, low-resolution stars; second row: the noiseless, highly resolved observations; third row: the WFE at the corresponding FOV positions.}
    \label{fi:observation_example}
\end{figure}

We generate a WFE field as in \citet{liaudat2023} using a maximum Zernike order of $45$, and a quadratic polynomial spatial variation per Zernike order. 
We enforce that the total amount of optical aberration at any position in the FOV ($\text{WFE}_{rms}$) is within expected limits.
A detailed list of parameter values is given in \autoref{tab:parameters} in the Appendix. 
We sample the GT WFE at random positions on the FOV and simulate in-focus star observations considering $13$ different spectral classes, as in \citet{liaudat2023}.
We reduced the number of spectral bins from $20$ to $8$ to minimise the computational cost of the numerical experiments. 
We simulate a total of $2\,400$ stars with a randomly drawn S/N ratio between $10$ and $110$, which we then split in $2\,000$ training stars and $400$ testing stars. 
The testing stars are generated without noise to provide a clearer study of the pixel-related metrics.

\Autoref{fi:observation_example} shows three examples of stellar observation simulations at different positions in the FOV. 
The first row shows noisy, low-resolution observations, the second row shows noiseless, highly resolved observations, and the third row shows the WFEs at the corresponding positions in the FOV. 
As in \citet{liaudat2023} we adopted certain simplifications such as not adding intra-pixel shifts to the postage stamps, not considering discontinuities in the WFE field, not including masked areas (e.g., due to cosmic rays or other phenomena), and an additive Gaussian noise assumption.

\subsection{Results}
%===================
\begin{table}
    \centering
    \caption{\small Results for the WaveDiff model including feature projection and NP model reset at the end of each cycle, as discussed in \autoref{sec:new_wfe_optim_algorithm}. The  proposed model, "Ours (Full)",  uses both parametric and NP parts for inference, while the "Ours (Param)" model uses only the parametric part. The results are compared to the original WaveDiff optimisation strategy \citep{liaudat2023}. The relative error is shown in the WFE-space, in the low-resolution pixel space (LR), and in the super-resolved pixel space (SR). For each metric and model, we show the median, standard deviation, and best of three realisations.}
    \resizebox{\columnwidth}{!}{%
    \begin{tabular}{@{}cccc@{}}
    \toprule
     & \multicolumn{3}{c}{Median $\pm$ Std Dev (best) {[}\%{]}}                     \\ \cmidrule(l){2-4} 
    Model      & WFE RMSE & LR Pixel RMSE & SR Pixel RMSE \\ \midrule
    WF Original    & $29.1 \pm 11.9$ (25.1) & $ 0.4 \pm 0.2 $ (0.4) & $ 0.9 \pm 0.3 $ (0.8) \\
    Ours (Full)   & $6.4 \pm 1.2 $ (6.1)   & $ \bm{0.3 \pm 0.1} $ (0.3) & $ \bm{0.7 \pm 0.2}$ (0.7)  \\
    Ours (Param) & $ \bm{3.4 \pm 0.9}$ (3.1)   & $ 0.4 \pm 0.2 $ (0.4) & $ \bm{0.7 \pm 0.2} $ (0.7) \\ \bottomrule
    \end{tabular}
    }
    \label{tab:results}
\end{table}
In this section, we present the results of the new optimisation framework that includes wavefront feature projections. 
We evaluate the ability of the model to retrieve the GT PSF field both in wavefront and pixel space. 
The latter is evaluated in the resolution of the observations (low-resolution, LR) and in super-resolution simulations (SR). 
The results are compared with the original WaveDiff optimisation strategy using \autoref{al:wavediff_training_original}, which does not include the wavefront projections and uses two cycles optimisation cycles. 
Note that although we use the original training algorithm, the new scenario in which the model is capable of representing the GT PSF field still holds. 
Additionally, we investigate the second order moment shape errors of the recovered PSFs, which are presented in \autoref{apx:shape_results}.

\Autoref{tab:results} summarises the WaveDiff PSF field recovery results with the new optimisation scenario\footnote{The jobs used to run these experiments are available at \href{https://github.com/tobias-liaudat/wf-psf/tree/v1.4.0/experiments/jz-phase-retrieval/jobs}{https://github.com/tobias-liaudat/wf-psf/tree/v1.4.0/experiments/jz-phase-retrieval/jobs}, where all the parameters and hyperparameters used are detailed.}. 
The first row corresponds to the baseline case without projections (WF Original). 
The second row shows the results obtained with the new model, using both parametric and non-parametric contributions. 
Finally, the third row presents the results of the new model, where only the parametric part is used for evaluation. 
The first column reports the relative mean square error (RMSE) of the wavefront recovery, the second column shows the low-resolution pixel RMSE, and the third column presents the super-resolution pixel RMSE.
The absolute WFE rms errors are shown in \autoref{apx:absolute_wfe_errors} both in nanometres and $\lambda^{-1}$. 
For each case, the table lists the median over three realisations, together with its standard deviation across all test FOV positions, and, in parentheses, the best-performing realisation.

\begin{figure}
    \centering
    \includegraphics[width=0.48\textwidth,trim={0cm .2cm 0cm 0.2cm},clip]{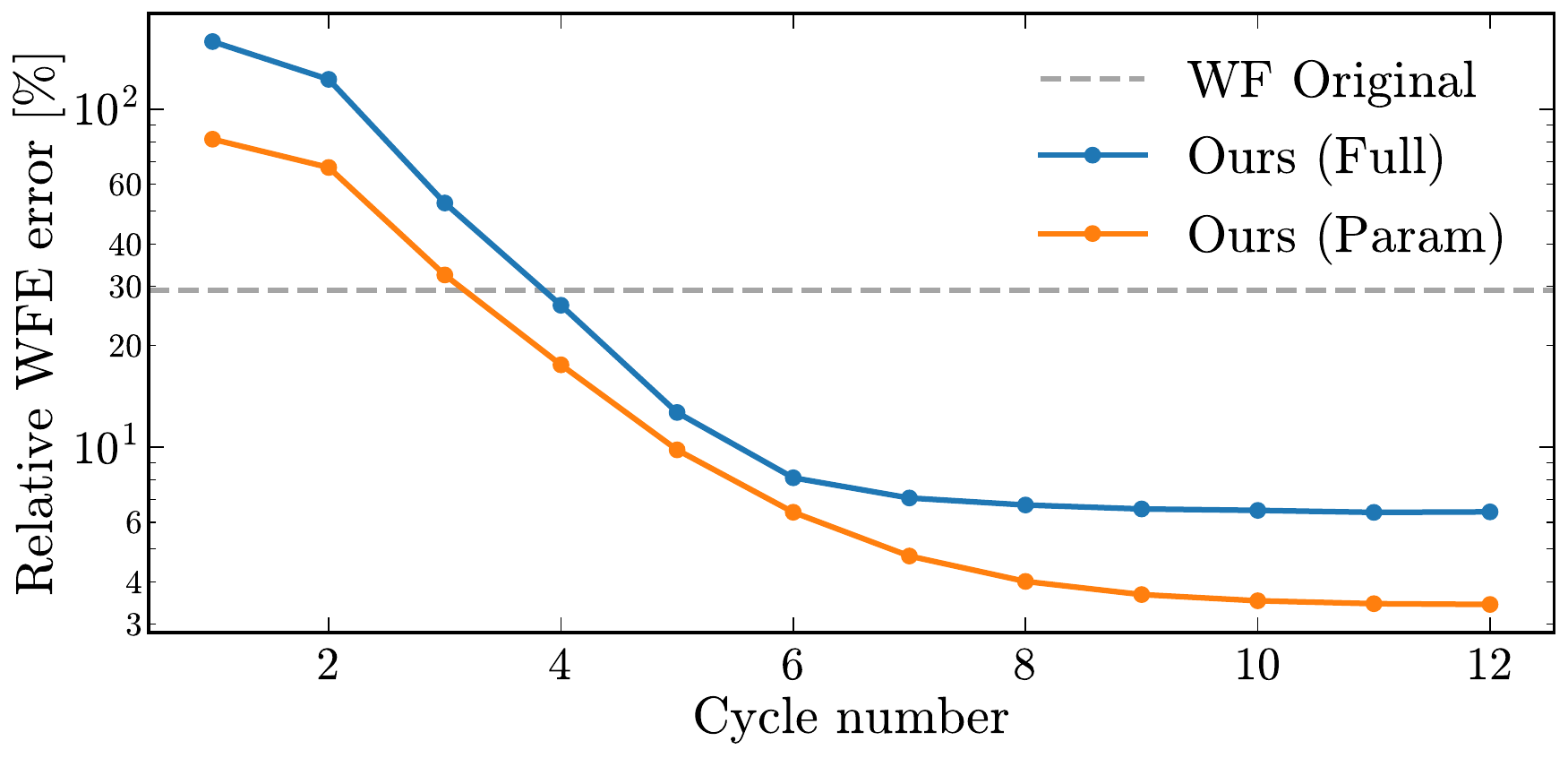}
    \caption{\small Relative WFE error for the new WaveDiff optimisation scenario for each optimisation cycle. 
    The "Ours (Full)" model uses both parametric and non-parametric parts for inference, while the "Ours (Param)" model uses only the parametric part. 
    The grey dashed line shows the results of the original optimisation scenario.}
    \label{fi:cycle_evol_rel_opd_error}
\end{figure}

\begin{figure}
    \centering
    \includegraphics[width=0.48\textwidth,trim={0cm .2cm 0cm 0.2cm},clip]{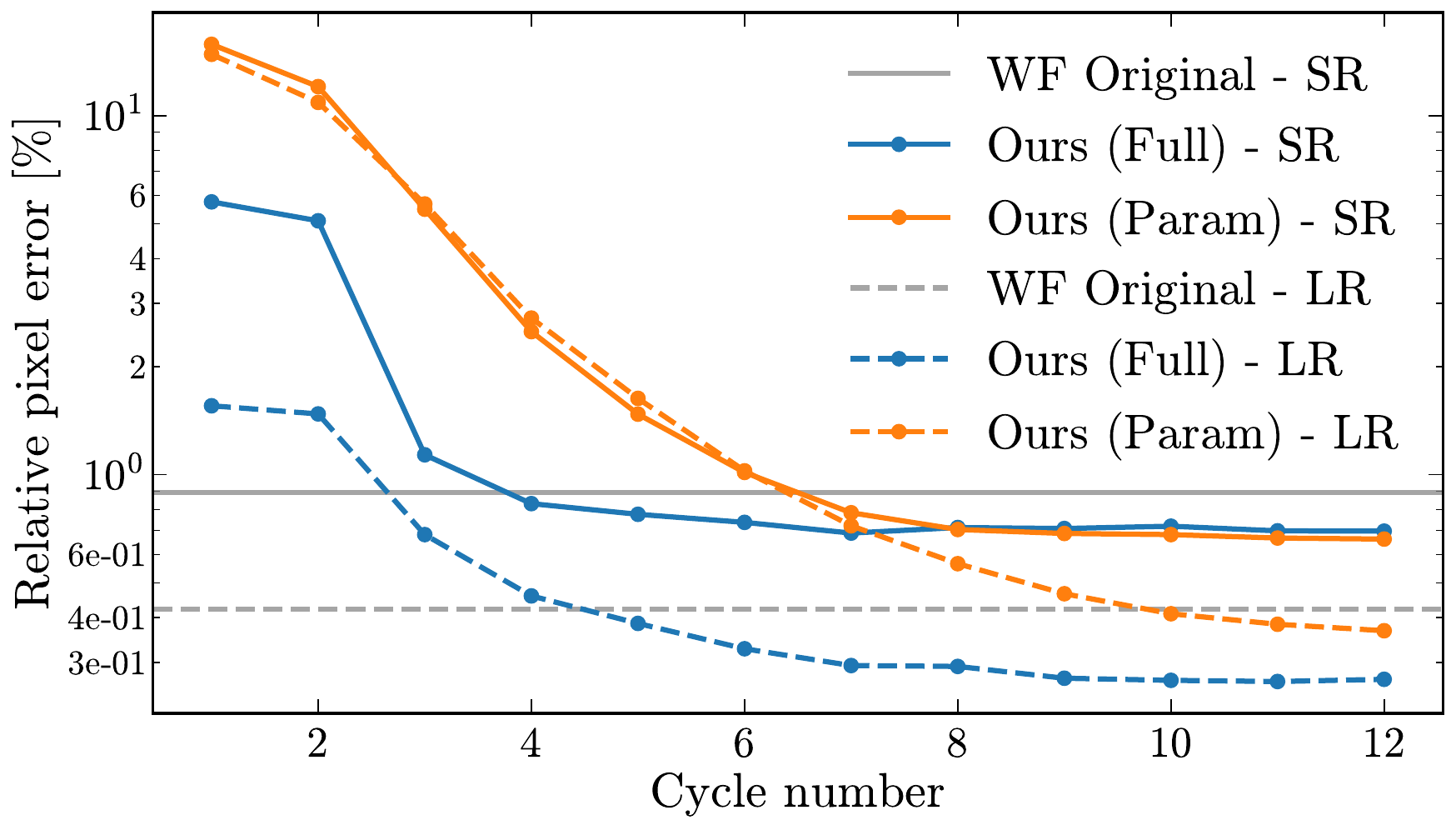}
    \caption{\small Relative pixel error at observation resolution (LR) and super-resolution (SR) across the 12 optimisation cycles. 
    The "Ours (Full)" model uses both parametric and non-parametric parts for inference, while the "Ours (Param)" model uses only the parametric part. 
    The grey lines show the results of the original optimisation scenario.}
    \label{fi:cycle_evol_rel_pixel_error_bothRes}
\end{figure}

The new optimisation scenario leads to a substantial reduction in the WFE error, from about $30\%$ to $3.4\%$, decreasing by nearly an order of magnitude compared to the original model. 
We also observe that the parametric-only model achieves better WFE performance than the full model, although the full model performs slightly better it in pixel space.
This behaviour is expected since the non-parametric model is optimised to decrease the pixel-space error, without physical restrictions over the learnt wavefront features.
\Autoref{fi:cycle_evol_rel_opd_error} presents the evolution of the relative WFE error over the $12$ optimisation cycles, where we can observe the convergence of the algorithm.
The pixel-space PSF recovery results of the new optimisation scenario remain consistent with those of the original WaveDiff model, with a slight improvement in both low and super resolution.
The evolution of the relative pixel error over the 12 cycles is shown in \autoref{fi:cycle_evol_rel_pixel_error_bothRes}.

\subsection{Single star phase retrieval}
%=======================================
To put the WaveDiff method in context in the problem of WFE phase retrieval, we conducted an experiment in which we attempted to recover the WFE of a single in-focus observation drawn from the simulation dataset presented in \autoref{sec:sims}. 
As mentioned in \autoref{sec:sota}, this is a genuinely complex and ill-posed problem, since we have only in-focus intensity measurements (PSF sample) for estimating the phase of a complex signal (the underlying WFE). Many phase retrieval methods make use of focus diversity to overcome this difficulty.
We implemented a simple WFE recovery method, based on WaveDiff's differentiable forward optics, which models the WFE as a single non-parametric feature. 
We estimate the WFE by solving \autoref{eq:optim_pixel} using a gradient descent algorithm.

\begin{figure}
    \centering
    \includegraphics[width=.9\columnwidth]{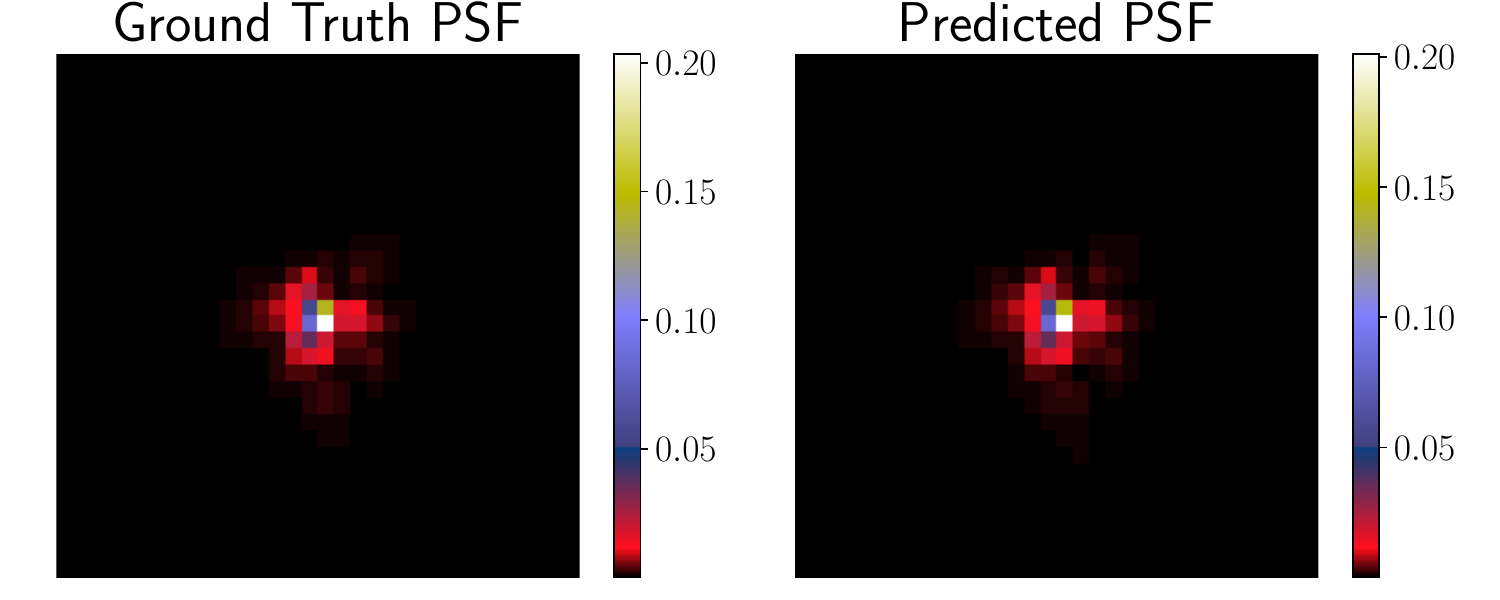}
    \caption{\small Ground truth and predicted PSF for a noiseless in-focus star.}
    \label{fi:single_star_PR_pixel}
\end{figure}

\begin{figure}
    \centering
    \includegraphics[width=0.9\columnwidth]{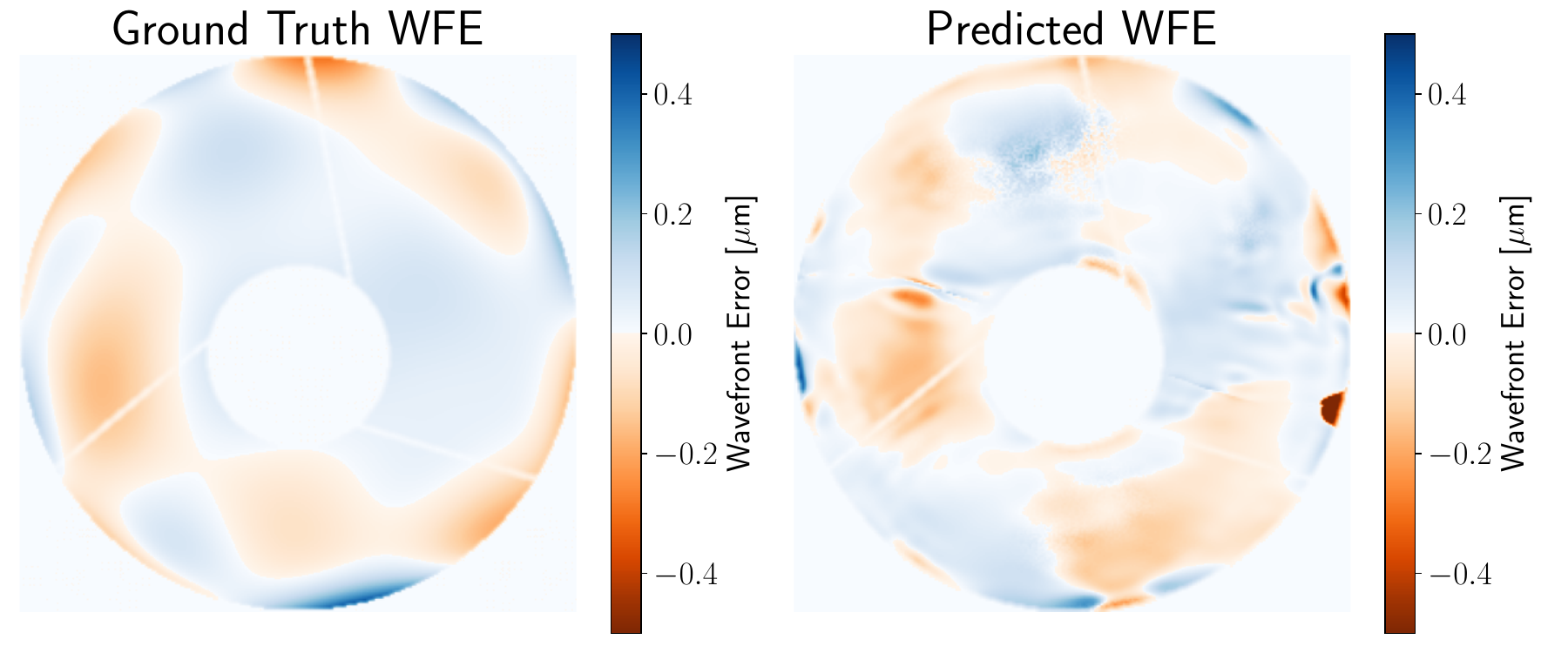}
    \caption{\small Ground truth and predicted WFE for a noiseless in-focus star.}
    \label{fi:single_star_PR_wavefront}
\end{figure}

We fit the single non-parametric wavefront feature using an Adam optimiser for $100$ epochs, with a learning rate of $0.03$. The recovered PSF and WFE are shown in \autoref{fi:single_star_PR_pixel} and \autoref{fi:single_star_PR_wavefront}, respectively, and are compared to the ground truth. 
The method achieves a pixel-space reconstruction error of $1.2\%$, consistent with the results presented above. 
However, in wavefront space the residual WFE rms is $73$ nm, that is $0.1 \; \lambda^{-1}$ (considering the central wavelenght $\lambda=725$ nm), giving an extremely high relative error, close to $100\%$. 
This outcome is expected, since a single in-focus stellar observation provides very limited phase information for recovering the GT WFE. 
Furthermore, no regularisation or physical constraints are imposed on the learned wavefront feature. 
In this test, we used a single noise-free star, as experiments with noisy stars resulted in pixel-space errors exceeding $10\%$.

The aim of this experiment is to illustrate the difficulty of estimating phase information from a single in-focus stellar observation. 
WaveDiff overcomes this limitation by leveraging observations of multiple stars across the FOV, jointly modelling the WFE field and exploiting the spatial diversity of the PSF field.

\subsection{Wavefront retrieval evolution}
To gain qualitative insight into the behaviour of the proposed WaveDiff optimisation process, we examine the evolution of the WFE for a representative star over the $12$ optimisation cycles. 
The selected target star, together with the corresponding WaveDiff super-resolved PSF reconstruction, is shown in \autoref{fi:pixel_PSF_example} in the Appendix. 
The ground-truth and reconstructed WFEs are presented in \autoref{fi:WFE_reconstruction}, along with their decomposition into parametric and non-parametric components.

\begin{figure*}
    \captionsetup[subfigure]{justification=centering}
    \centering
    \begin{subfigure}[b]{\textwidth}
        \centering
        \includegraphics[width=.67\linewidth]{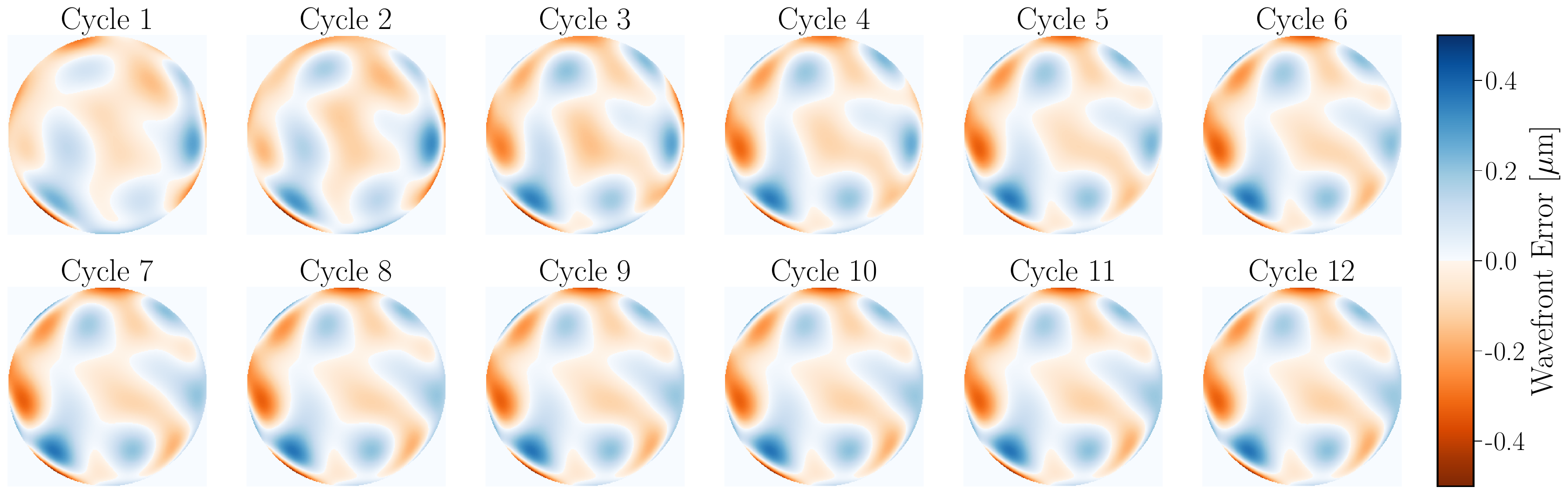}
        \caption{Parametric part WFE evolution at the end of each cycle.}
        \label{fi:WFE_evolution_a}
    \end{subfigure}\\
    \begin{subfigure}[b]{\textwidth}
        \centering
        \includegraphics[width=.67\linewidth]{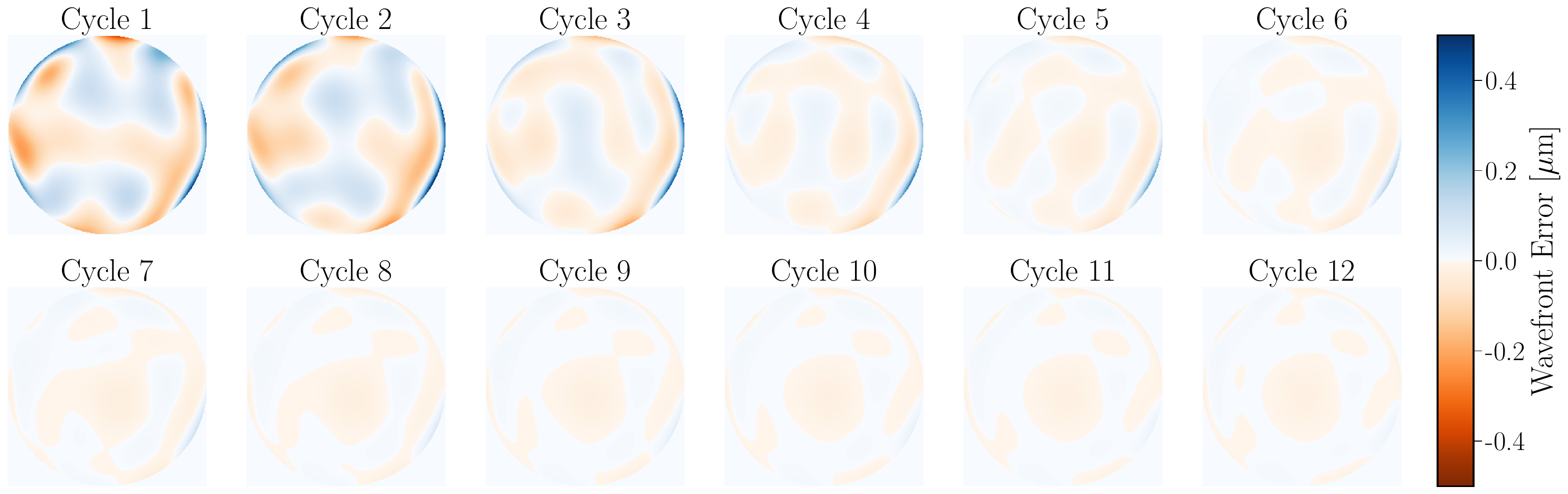}
        \caption{Evolution of the parametric part WFE residual at the end of each cycle with respect to the GT WFE.}
        \label{fi:WFE_evolution_b}
    \end{subfigure}\\
    \begin{subfigure}[b]{\textwidth}
        \centering
        \includegraphics[width=.67\linewidth]{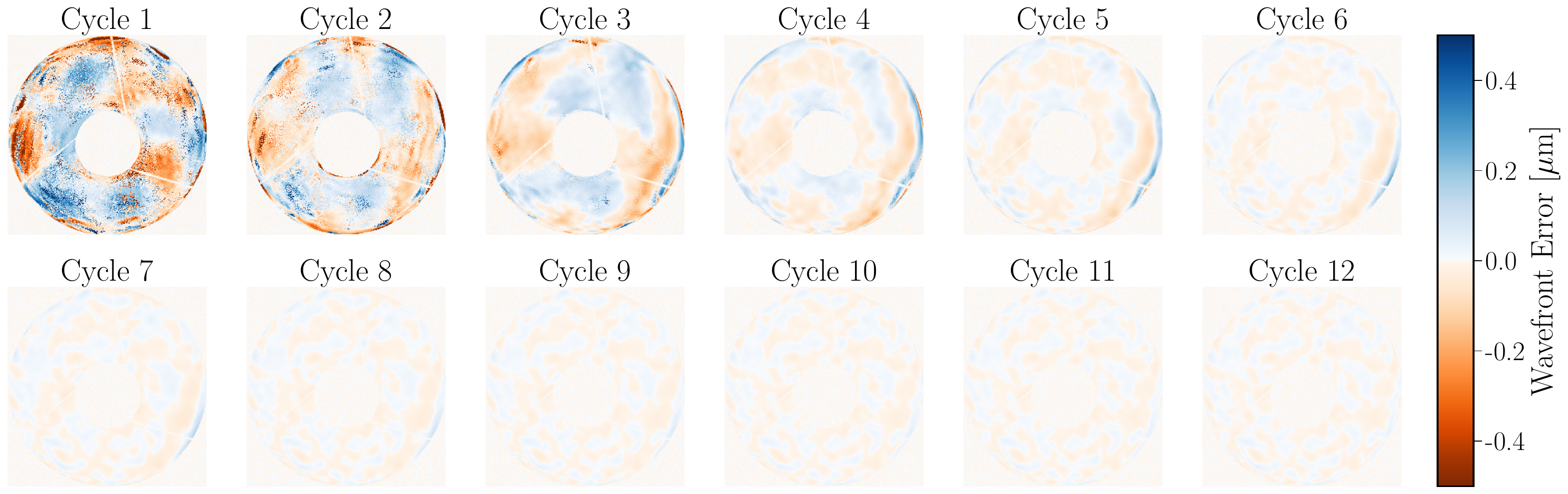}
        \caption{Evolution of the non-parametric part before being projected at the end of each cycle.}
        \label{fi:WFE_evolution_c}
    \end{subfigure}
    \caption{\small Evolution of different WFE parts of the model as the number cycle increases. The same scale is used for all the images in each figure.}
    \label{fi:WFE_evolution}
\end{figure*}
\Autoref{fi:WFE_evolution} shows the evolution of the WFE for the two components of the model across the optimisation cycles. 
\Autoref{fi:WFE_evolution_a} illustrates the evolution of the parametric component, highlighting how it progressively approaches the ground-truth WFE after each projection (see also \autoref{fi:WFE_reconstruction_f}). 
\Autoref{fi:WFE_evolution_b} displays the residual of the parametric component with respect to the ground-truth WFE, which steadily decreases in amplitude over successive cycles.
Finally, \autoref{fi:WFE_evolution_c} presents the non-parametric contribution at the end of each cycle, prior to its projection onto the parametric model.
As the cycles progress, the magnitude of this contribution decreases, indicating that the non-parametric model has progressively less to correct and giving a clear sign of convergence. 
Moreover, the non-parametric contribution closely reproduces the residual patterns of the parametric component from the preceding cycle, demonstrating that the non-parametric model effectively compensates for the parametric WFE error. 
This behaviour is enabled by reinitialising the non-parametric model after each projection, which prevents it from remaining in local minima and allows further optimisation.

%=======================================================================================
\section{Conclusions}
\label{sec:conclusion}
%=======================================================================================
In this paper, we have presented a new optimisation framework for the WaveDiff model, which represents the PSF in wavefront space. 
The proposed approach integrates a wavefront feature projection algorithm that facilitates the interaction between the two components of this model: the parametric and the non-parametric contributions. 
Since optimising the parametric component of WaveDiff is particularly challenging, especially when using such a high order of Zernike polynomials ($n_{\text{Z}}=45$), we focus exclusively on optimising the non-parametric component and use wavefront projection to transfer the information learned by this component to the parametric one. 
This approach allows us to more effectively exploit the spatial diversity of the optical aberrations to recover the WFE field.

This work addresses a phase retrieval problem in which we aim to estimate the WFE from in-focus, intensity-only observations. Unlike conventional phase retrieval, which typically recovers the WFE at a single field position from multiple defocused observations of one star, we have only one in-focus intensity measurement per FOV position, each corresponding to a distinct WFE. Nevertheless, all stars share a common underlying WFE field, modelled by a quadratic spatial polynomial for each Zernike order, and recovery is made tractable by simultaneously exploiting this shared structure and the spatial diversity of optical aberrations across the field of view.

We conducted numerical experiments to validate the proposed optimisation framework. 
The results are notable in that the method accurately reconstructs the underlying ground-truth WFE without any prior knowledge of the true WFE, making it the first demonstrated method to address in a single framework the wide-field WFE recovery from noisy in-focus-only polychromatic observations exploiting non-parametric wavefront features.
Accurate WFE recovery is particularly valuable, as it enables the construction of an accurate PSF model and provides diagnostic insight into the telescope's optical system, allowing the identification of potential anomalies.
The proposed approach achieves a relative WFE reconstruction error of $3.4\%$, which is an order of magnitude lower than that obtained with the previous WaveDiff optimisation procedure described in \citet{liaudat2023}. 
In addition, it improves the pixel-space reconstruction by reducing the relative PSF recovery error by $20\%$. 

Although in this study the parametric model is capable of representing the WFE GT, and this is why the parametric evaluation outperforms the full semi-parametric evaluation in the WFE space, WaveDiff remains a semi-parametric data-driven model. 
Thus, we can imagine more complex scenarios in which we use the new approach to estimate the base parametric part and still use the non-parametric part for higher-order components that cannot be captured by the parametric model. 
In future work, we plan to explore the proposed methodology in more realistic scenarios. 
In particular, to asses the robustness to further realistic noise, instrumental artifacts, and uncertainties in the input stellar SEDs, which were assumed to be fully known for this study.

%=======================================================================================
% \section*{Data Availability}
\small
\begin{dataavailability}
The scripts and notebooks for reproducing the results and figures in this paper are available at \href{https://github.com/tobias-liaudat/wf-psf/tree/v1.4.0}{https://github.com/tobias-liaudat/wf-psf/tree/v1.4.0}. 
The trained models and metrics are available at \href{https://zenodo.org/records/17590010}{https://zenodo.org/records/17952609}.
\end{dataavailability}
%=======================================================================================

%=======================================================================================
\begin{acknowledgements}
%=======================================================================================
This work was granted access to the HPC resources
of IDRIS under the allocation 2024-AD011012983R3 and 2025-AD011012983R4 made by GENCI. 
A preliminary part of this work was presented in the PhD thesis \citet{liaudat2022}.
This work was supported by the TITAN ERA Chair project (contract no. 101086741) within the Horizon Europe Framework Program of the European Commission, and the Agence Nationale de la Recherche (ANR-22- CE31-0014-01 TOSCA).
\end{acknowledgements}
\vspace{-1cm}

%=======================================================================================
% \bibliographystyle{rasti}
\bibliographystyle{aa}
\bibliography{bib_phase_retrieval}
\normalsize
\clearpage
%=======================================================================================
\begin{appendix}
\onecolumn

\section{Notation}
The following table summarises the notation adopted in this article with regard to PSF modelling in wavefront space. 
We adopt the notation defined in \citet{liaudat2023_2}.
\FloatBarrier
% \begin{table}
    \begin{center}
    \captionof{table}{Coordinates and notation used throughout this article.}
    \label{tb:variable}
        \begin{tabular}{cl} 
        \toprule
        Variable & \multicolumn{1}{c}{Description}   \\
        \midrule
        \multicolumn{2}{c}{\textit{Coordinates}} \\
        \midrule
        $(x,y)$                 & Pupil plane coordinates \\
        $(u,v)$                 & Image or focal plane coordinates \\
        % $(\xi, \eta)$           & Object plane coordinates \\
        $(\bar{u}, \bar{v})$, $(\bar{x},\bar{y})$    & The discrete counterpart of the focal and pupil plane coordinates \\
        % $\mathbf{p}_i$          & 3D spatial coordinate \tl{see if we can remove this}  \\
        $\lambda$               & Wavelength  \\
        $t$                     & Time \  \\
        \midrule
        \multicolumn{2}{c}{\textit{Notation}} \\
        \midrule
        $\mathcal{I}, \mathcal{H}, \ldots$  & Calligraphic uppercase variables are continuous functions \\
        $I, H, \ldots$                      & Uppercase variables are matrices \\
        $c_m, b_{1}^{k}, \ldots$            & Lowercase variables are scalars \\
        $I_\mathrm{img}(\bar{u},\bar{v};t|u_i,v_i) \in \mathbb{R}$ & Pixel value at position $(\bar{u},\bar{v})$ for the image $I_\mathrm{img}$ with its \\
        & centroid at position $(u_i,v_i)$ observed at time $t$. \\
        $I_{\mathrm{img},} (\cdot|u_i,v_i) \in \mathbb{R}^{p \times p}$ & Observed image  with its centroid at position $(u_i,v_i)$  \\
        \bottomrule
        \end{tabular}
    \end{center}
% \end{table}
\FloatBarrier

\section{Projected non-parametric subtraction}
\label{apx:projection}
In \autoref{sec:non-param_projection} we described how to modify the parametric part to account for the information learned by the NP part. 
We then need to modify the NP part so that the information that has just been transferred to the parametric part is not replicated in the NP contribution. 
The sum of both parts of the model should remain constant after the projection.

We rewrite the NP part of the model from \autoref{eq:dd_wfe_model} using a tensor product as follows
\begin{equation}
    \Phi^{\text{NP}} = \sum_{k=1}^{n_{\text{NP}}} \underbrace{\bm{\pi}^{\text{NP}^{\text{T}}} \, \bm{a}_{k}}_{\in \mathbb{R}} \, S_{k}^{\text{NP}} = \bm{\pi}^{\text{NP}^{\text{T}}} \, A \, \bm{S}^{\text{NP}} \, ,
    \label{eq:dd_wfe_model_tensors}
\end{equation}
where $\bm{S}^{\text{NP}} \in \mathbb{R}^{n_{\text{NP}} \times K \times K}$ is a tensor containing all the learned NP features $S_{i}^{\text{NP}} \in \mathbb{R}^{K \times K}$.
In order to express what was transmitted to the parametric model, we will reformulate the $S$ tensor in terms of the projection onto the first $n_{\text{Z}}$ Zernike polynomials of the NP features.
Then, we remove the Zernike projections from each NP feature. 
However, we would be neglecting that only the WFE projection associated with FOV spatial variations, i.e. polynomial in $u$ and $v$, of order less or equal to $d_{\text{Z}}$ was transmitted. 
The polynomial degree of the NP part is in general bigger, $d_{\text{NP}} > d_{\text{Z}}$. Consequently, the higher-frequency FOV spatial variations are not transmitted as the parametric part cannot represent them. 
The solution is straightforward if we suppose the $A$ matrix is the identity matrix. 
In that case, we have to remove the projection over the first $n_{\text{Z}}$ Zernike polynomials for the first $n_{\text{d}_{\text{Z}}}$ matrices of $\bm{S}^{\text{NP}}$. 
However, after the optimisation of the model, the mixing matrix $A$ will not be the identity, and it will mix the NP features $S^{\text{NP}}_{k}$ as follows
\begin{equation}
    \bm{\Tilde{S}}^{\text{NP}} = A \, \bm{S}^{\text{NP}} = \sum_{i=1}^{n_{\text{NP}}} \sum_{k=1}^{n_{\text{NP}}} \bm{a}_{k}[i] S_k^{\text{NP}} .
\end{equation}

The mixed NP feature tensor $\bm{\Tilde{S}}^{\text{NP}}$ has a direct matching with the spatial variations from $\bm{\pi}^{\text{NP}}$. 
Therefore, we remove the contribution of the first $n_\text{Z}$ Zernike polynomials to the first $n_{\text{d}_{\text{Z}}}$ matrices of the mixed NP features $\bm{\Tilde{S}}^{\text{NP}}$. 
The only parameters we can modify from the NP part is the $\bm{S}^{\text{NP}}$ tensor. 
The vector $\bm{\pi}^{\text{NP}}$ and the matrix $A$ are shared by NP features of a higher FOV spatial variation, and therefore we should not modify them. 
We look for a new target tensor $\bm{S}^{{*}^{\text{NP}}}$ such that once it is multiplied by $A$, its first $n_{d_{\text{Z}}}$ matrices are orthogonal to the first $n_{\text{Z}}$ Zernike polynomials. 
We denote $\bm{\Tilde{S}}^{{*}^{\text{NP}}} = A\, \bm{S}^{{*}^{\text{NP}}}$, the mixed target tensor. 
This new tensor, $\bm{\Tilde{S}}^{{*}^{\text{NP}}}$, can be computed by subtracting the Zernike polynomial projection over the first $n_{d_{\text{Z}}}$ matrices as follows
\begin{equation}
    \Tilde{S}^{{*}^{\text{NP}}}_{k} = \begin{dcases}
    \Tilde{S}_{k}^{\text{NP}} - \sum_{i=1}^{n_{\text{Z}}} \text{P}_{S_{i}^{\text{Z}}}\big( \Tilde{S}_{k}^{\text{NP}} \big) \, S_{i}^{\text{Z}}  \quad &\text{if} \;\; 1 \leq k \leq n_{\text{d}_{\text{Z}}} \\
    \\
    \Tilde{S}_{k}^{\text{NP}} \quad &\text{if} \;\; n_{\text{d}_{\text{Z}}} < k \leq n_{\text{NP}}\\
    \end{dcases} \,.
\end{equation}

Once we have computed the tensor $\bm{\Tilde{S}}^{{*}^{\text{NP}}}$, we can recover the desired parameters by unmixing the tensor as follows
\begin{equation}
    \bm{S}^{{*}^{\text{NP}}} = A^{-1} \,\bm{\Tilde{S}}^{{*}^{\text{NP}}} \,.
\end{equation}

We cannot guarantee that once the optimisation algorithm has converged, the matrix $A$ is nonsingular. 
However, in practice, the matrix is initialised with the identity and converges to a nonsingular matrix. 

\section{WaveDiff Training}
\label{apx:algorithms}
\FloatBarrier
In this appendix, we present the training algorithms for the WaveDiff model. 
First, we show the original training procedure \citep{liaudat2023} and then the proposed adaptation.

% Original Training
\begin{algorithm*}
    \DontPrintSemicolon
    \caption{Original WaveDiff training procedure}
    \label{al:wavediff_training_original}
    \SetInd{0.2em}{2em} % Control indentation block width
    \textbf{Hyperparameters}: \\
    $\{\eta_{m}^{\text{NP}} \}_{m=1, \ldots, m_{\max}}$, 
    $\{\eta_{m}^{\text{Z}} \}_{m=1, \ldots, m_{\max}}$ \tcp*{Learning rates}
    $\{N_{\text{ep}, m}^{\text{NP}} \}_{m=1, \ldots, m_{\max}}$,
    $\{N_{\text{ep}, m}^{\text{Z}} \}_{m=1, \ldots, m_{\max}}$\tcp*{Number of epochs}
    \bigskip
    
    \textbf{Initialization}: \\  
    $S_{k}^{\text{NP}} \sim \mathcal{U}[-10^{-3}, 10^{-3}]^{K \times K}, \forall k$ \tcp*{Non-parametric features}
    $A_{\text{mix}} \gets I_{n_{\text{NP}}}$ \tcp*{Non-parametric mixing matrix}
    $\pi^{\text{NP}}_{[l,m]} \sim \mathcal{U}[-10^{-2}, 10^{-2}], \forall \, l+m \leq d_{\text{NP}}$ \tcp*{Non-parametric feature weights}
    $\pi^{\text{Z}}_{l,[s,t]} \sim \mathcal{U}[-10^{-2}, 10^{-2}], \forall \, l \leq n_{\text{Z}},\, s+t \leq d_{\text{Z}}$ \tcp*{Parametric feature weight}
    $\hat{\sigma}_i \leftarrow \text{MAD noise estimation} (\bar{I}(u_i, v_i))$ \tcp*{Estimate noise level}
    Generate Zernike polynomial maps $S_{l}^{\text{Z}}, \forall \, l \leq n_{\text{Z}},$ and telescope obscuration $\mathcal{P}$ \\
    
    \bigskip
    \textbf{Optimisation procedure}: \\
    \For{$m=1$ to $m_{\max}$}{
            Solve \autoref{eq:optim_pixel} for $\pi^{\text{Z}}_{l,[s,t]}$, \tcp*{Optimise parametric part}
            $\quad\quad\quad\quad\quad \forall \, l \leq n_{\text{Z}},\, s+t \leq d_{\text{Z}}$ using $\eta_{m}^{\text{Z}}$, $N_{\text{ep}, m}^{\text{Z}}$\\
            Solve \autoref{eq:optim_pixel} for $\pi^{\text{NP}}_{[l,m]},\, A_{\text{mix}},\, S_{k}^{\text{NP}}$, \tcp*{Optimise non-parametric part}
            $\quad\quad\quad\quad\quad \forall \, k \leq n_{\text{NP}},\, l+m \leq d_{\text{NP}}$ using $\eta_{m}^{\text{NP}}$, $N_{\text{ep}, m}^{\text{NP}}$\\
            
    }
    \bigskip
    Return PSF model: $\Phi_{\theta}$
    \vspace{0.05in}
\end{algorithm*}

% Proposed Training
\begin{algorithm*}
    \DontPrintSemicolon
    \caption{Proposed WaveDiff training procedure}
    \label{al:wavediff_training}
    \SetInd{0.2em}{2em} % Control indentation block width
    \textbf{Hyperparameters}: $\{\eta_{m}^{\text{NP}} \}_{m=1, \ldots, m_{\max}}$, $\{N_{\text{ep}, m}^{\text{NP}} \}_{m=1, \ldots, m_{\max}}$ \tcp*{Learning rates, Number of epochs}
    \bigskip
    
    \textbf{Initialization}: \\  
    $S_{k}^{\text{NP}} \sim \mathcal{U}[-10^{-3}, 10^{-3}]^{K \times K}, \forall k$ \tcp*{Non-parametric features}
    $A_{\text{mix}} \gets I_{n_{\text{NP}}}$ \tcp*{Non-parametric mixing matrix}
    $\pi^{\text{NP}}_{[l,m]} \sim \mathcal{U}[-10^{-2}, 10^{-2}], \forall \, l+m \leq d_{\text{NP}}$ \tcp*{Non-parametric feature weights}
    % $\pi^{\text{Z}}_{l,[s,t]} \sim \mathcal{U}[-10^{-2}, 10^{-2}], \forall \, l \leq n_{\text{Z}},\, s+t \leq d_{\text{Z}}$ \tcp*{Parametric feature weight}
    $\pi^{\text{Z}}_{l,[s,t]} \gets \mathbf0, \forall \, l \leq n_{\text{Z}},\, s+t \leq d_{\text{Z}}$ \tcp*{Parametric feature weight}
    
    $\hat{\sigma}_i \leftarrow \text{MAD noise estimation} (\bar{I}(u_i, v_i))$ \tcp*{Estimate noise level}
    Generate Zernike polynomial maps $S_{l}^{\text{Z}}, \forall \, l \leq n_{\text{Z}},$ and telescope obscuration $\mathcal{P}$ \\
    
    \bigskip
    \textbf{Optimisation procedure}: \\
    \For{$m=1$ to $m_{\max}$}{
            % Solve \autoref{eq:optim_problem} for $\pi^{\text{Z}}_{l,[s,t]}$, \tcp*{parametric part}
            % $\quad\quad\quad\quad\quad \forall \, l \leq n_{\text{Z}},\, s+t \leq d_{\text{Z}}$ using $\eta_{m}^{\text{Z}}$, $N_{\text{ep}, m}^{\text{Z}}$\\
            Solve \autoref{eq:optim_pixel} for $\pi^{\text{NP}}_{[l,m]},\, A_{\text{mix}},\, S_{k}^{\text{NP}}$, \tcp*{Optimise non-parametric part}
            $\quad\quad\quad\quad\quad \forall \, k \leq n_{\text{NP}},\, l+m \leq d_{\text{NP}}$ using $\eta_{m}^{\text{NP}}$, $N_{\text{ep}, m}^{\text{NP}}$\\
            Project non-parametric model using \autoref{al:projection_alg}\\
            Add projection result to the parametric part following \autoref{eq:parametric_update} \\
            Reset non-parametric part: $\pi^{\text{NP}}_{[l,m]} \sim \mathcal{U}[-10^{-2}, 10^{-2}], \forall \, l+m \leq d_{\text{NP}}$ \\
    }
    \bigskip
    Return PSF model: $\Phi_{\theta}$
    \vspace{0.05in}
\end{algorithm*}
\FloatBarrier
\clearpage
\section{Simulation parameters}
\label{apx:parameters}
We present the WaveDiff simulation parameters used in this study in \autoref{tab:parameters}.
\FloatBarrier

\begin{table}[h]
    \centering
    \small
    \caption{WaveDiff simulation parameters.}
    % \resizebox{\columnwidth}{!}{%
    \begin{tabular}{|l|l|l|}
    \hline
    \rowcolor[HTML]{E1E1E1} 
    Parameter                      & Description                & Value           \\ \hline
    $n_\text{Z}$                          & Maximum Zernike order      & $45$            \\ \hline
    $d_{\text{max}}$               &$C_k(x,y)$ polynomial degree& $2$             \\ \hline
    $n_\lambda$                     & Number of spectral bins    & $8$             \\ \hline
    $N_{\text{pix}}$               & Observation postage stamp size& $32$ px         \\ \hline
    $\mathcal{C}_{\rm MK}$                            & Number of stellar classes  & $13$            \\ \hline
    $S/N$                          & Signal-to-noise ratio range& $[10-110]$      \\ \hline
    $\text{WFE}_{rms}$                      & Maximum WFE RMS value      & $100$ nm         \\ \hline
    \end{tabular}
    \vspace{0cm}
    % }
    \label{tab:parameters}
\end{table}

\section{WaveDiff solution}
\begin{figure*}[h]
    \captionsetup[subfigure]{justification=centering}
    \centering
    \begin{subfigure}[b]{0.25\textwidth}
        \centering
        \includegraphics[width=\linewidth]{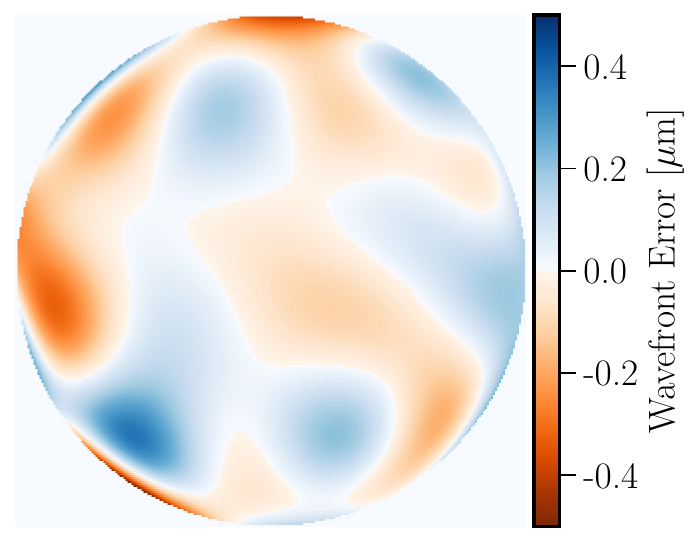}
        \caption{}
        \label{fi:WFE_reconstruction_a}
    \end{subfigure}
    \begin{subfigure}[b]{0.25\textwidth}
        \centering
        \includegraphics[width=\linewidth]{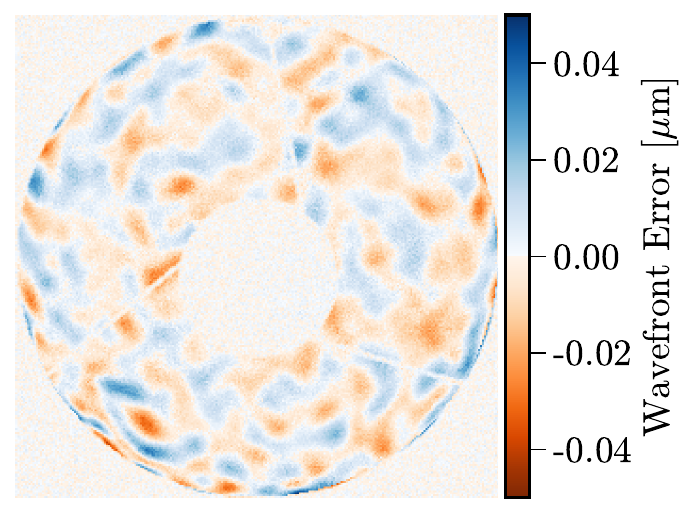}
        \caption{}
        \label{fi:WFE_reconstruction_b}
    \end{subfigure}
    \begin{subfigure}[b]{0.25\textwidth}
        \centering
        \includegraphics[width=\linewidth]{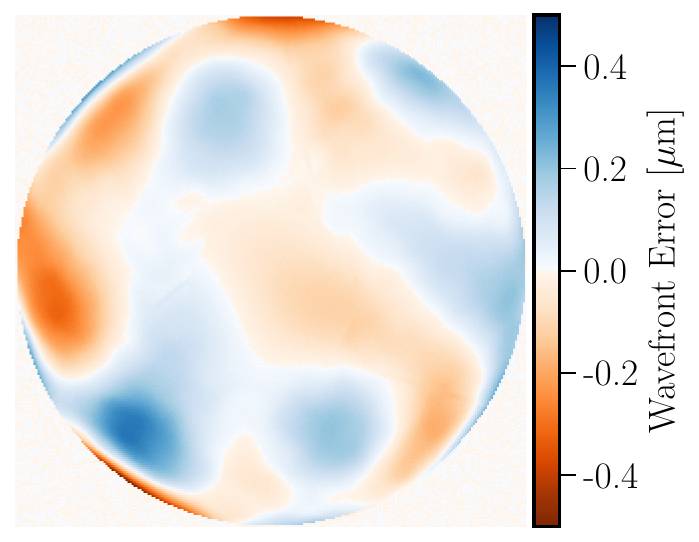}
        \caption{}
        \label{fi:WFE_reconstruction_c}
    \end{subfigure}\\
    \begin{subfigure}[b]{0.25\textwidth}
        \centering
        \includegraphics[width=\linewidth]{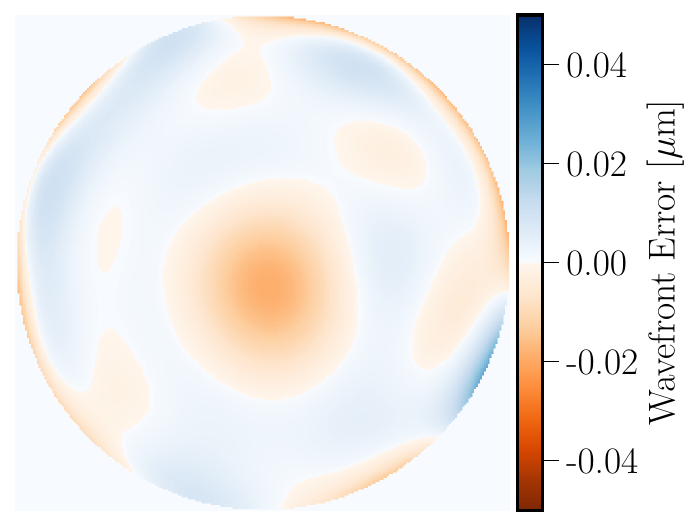}
        \caption{}
        \label{fi:WFE_reconstruction_d}
    \end{subfigure}
    \begin{subfigure}[b]{0.25\textwidth}
        \centering
        \includegraphics[width=\linewidth]{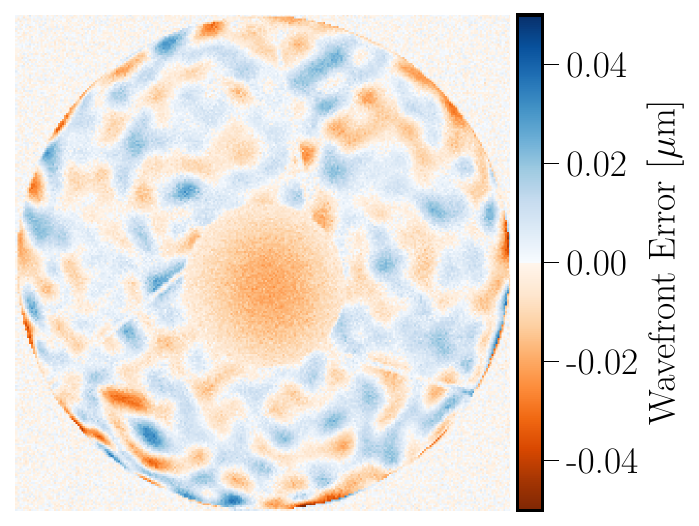}
        \caption{}
        \label{fi:WFE_reconstruction_e}
    \end{subfigure}
    \begin{subfigure}[b]{0.25\textwidth}
        \centering
        \includegraphics[width=\linewidth]{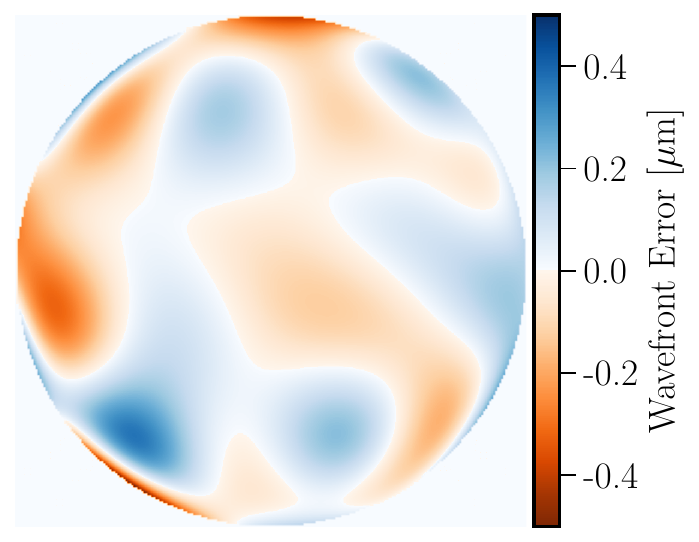}
        \caption{}
        \label{fi:WFE_reconstruction_f}
    \end{subfigure}
    \caption{WFE representations for a given training point in the FOV. The first row shows, from left to right, the parametric, NP, and the full WFE prediction for the new optimisation algorithm. The second row shows the residual WFE for the parametric and the full WFEs in the first and second columns, respectively, with respect to the ground truth WFE that is placed in the third column. 
    }
    \label{fi:WFE_reconstruction}
\end{figure*}

\begin{figure*}
    \captionsetup[subfigure]{justification=centering}
    \centering
    \begin{subfigure}[t]{0.22\textwidth}
        \centering
        \includegraphics[width=\linewidth]{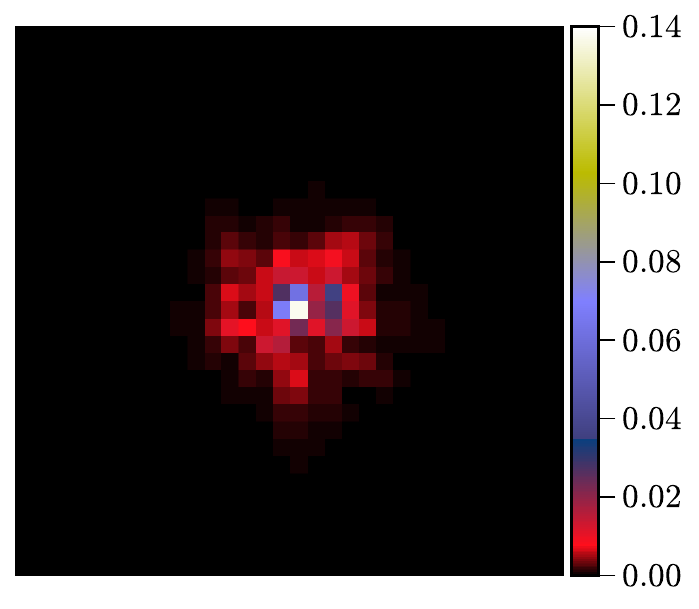}
        \caption{}
        \label{fi:pixel_PSF_example_a}
    \end{subfigure}
    \begin{subfigure}[t]{0.22\textwidth}
        \centering
        \includegraphics[width=\linewidth]{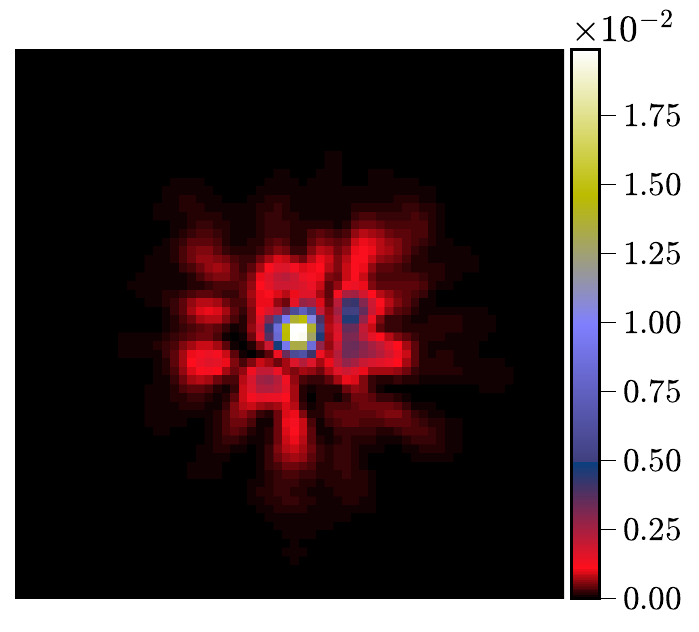}
        \caption{}
        \label{fi:pixel_PSF_example_b}
    \end{subfigure}
    \begin{subfigure}[t]{0.22\textwidth}
        \centering
        \includegraphics[width=\linewidth]{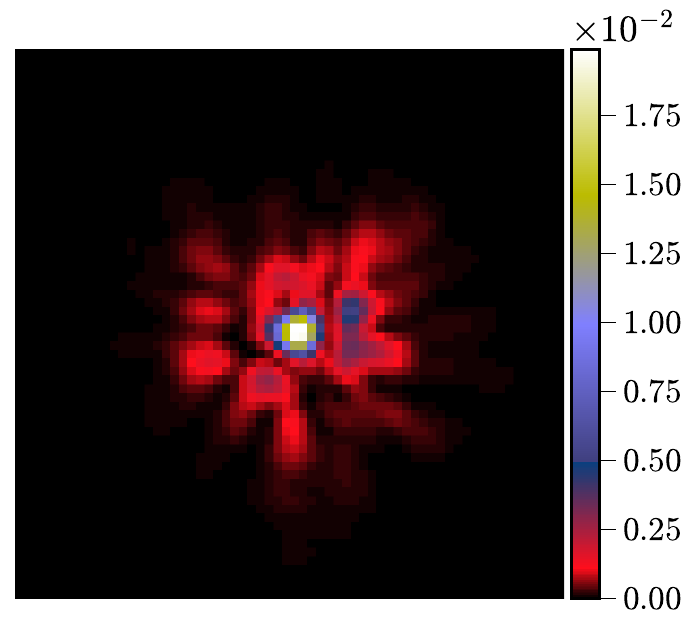}
        \caption{}
        \label{fi:pixel_PSF_example_c}
    \end{subfigure}
    \begin{subfigure}[t]{0.22\textwidth}
        \centering
        \includegraphics[width=\linewidth]{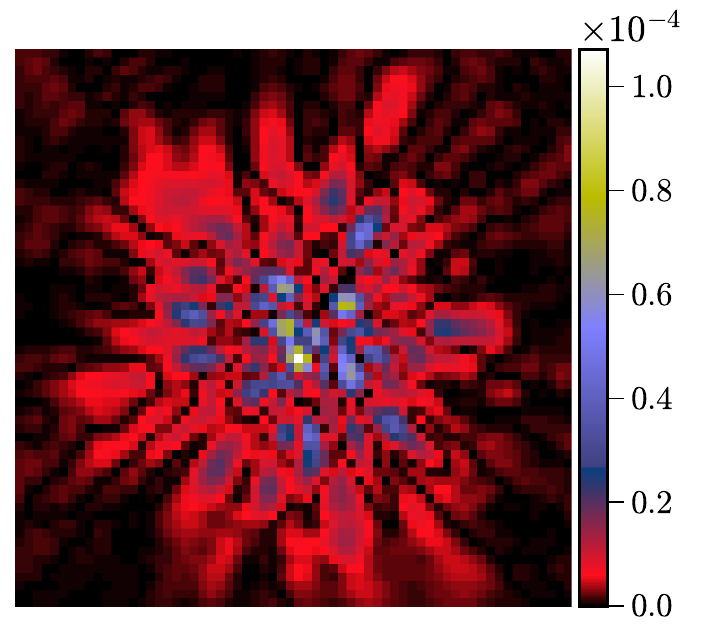}
        \caption{}
        \label{fi:pixel_PSF_example_d}
    \end{subfigure}
    \caption{Reconstructed PSF in pixel space for a given training point in the FOV. (a) GT PSF at observation resolution. (b) GT PSF at high resolution. (c) PSF reconstruction at high resolution. (d) Reconstruction residual (absolute value).}
    \label{fi:pixel_PSF_example}
\end{figure*}

This section presents the WaveDiff solution, both in wavefront space (\autoref{fi:WFE_reconstruction}) and pixel space (\autoref{fi:pixel_PSF_example}), for a test star, using the proposed optimisation strategy.

\FloatBarrier

\section{Shape results}
\label{apx:shape_results}
Given that PSF requirements in weak gravitational lensing surveys are typically expressed in terms of PSF ellipticity \citep{massey2012, cropper2013}, we present here the results of PSF shape recovery. 
For this purpose, we use a set of second order moment-based shape metrics, namely the ellipticity ($e_1$, $e_2$) and PSF size ($R^2$), which are defined in detail in \citet{liaudat2023}. 
The GT and predicted PSF moments are computed using the adaptive moment algorithm \citep{mandelbaum2005, hirata2003}, implemented in the HSM module of \texttt{Galsim} \citep{rowe2015}, with the super-resolved stars. 
The shape recovery results are summarised in \autoref{tab:shape_results}, which reports the RMSE over the two ellipticity components and the PSF size for both the original and new optimisation scenarios. 

\begin{figure*}[h]
    \captionsetup[subfigure]{justification=centering}
    \centering
    \begin{subfigure}[b]{\textwidth}
        \centering
        \includegraphics[width=0.4\textwidth]{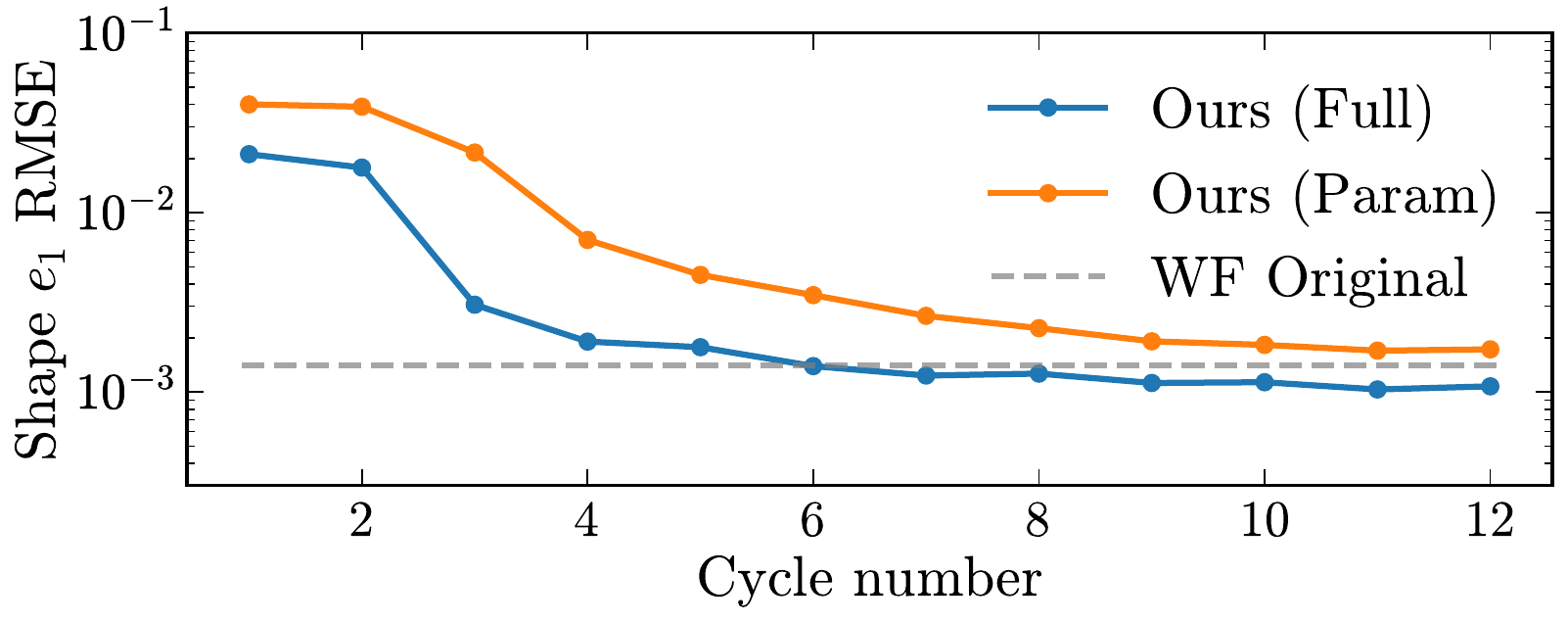}
        \caption{}
        \label{fi:e1_results}
    \end{subfigure}\\
    \begin{subfigure}[b]{\textwidth}
        \centering
        \includegraphics[width=0.4\textwidth]{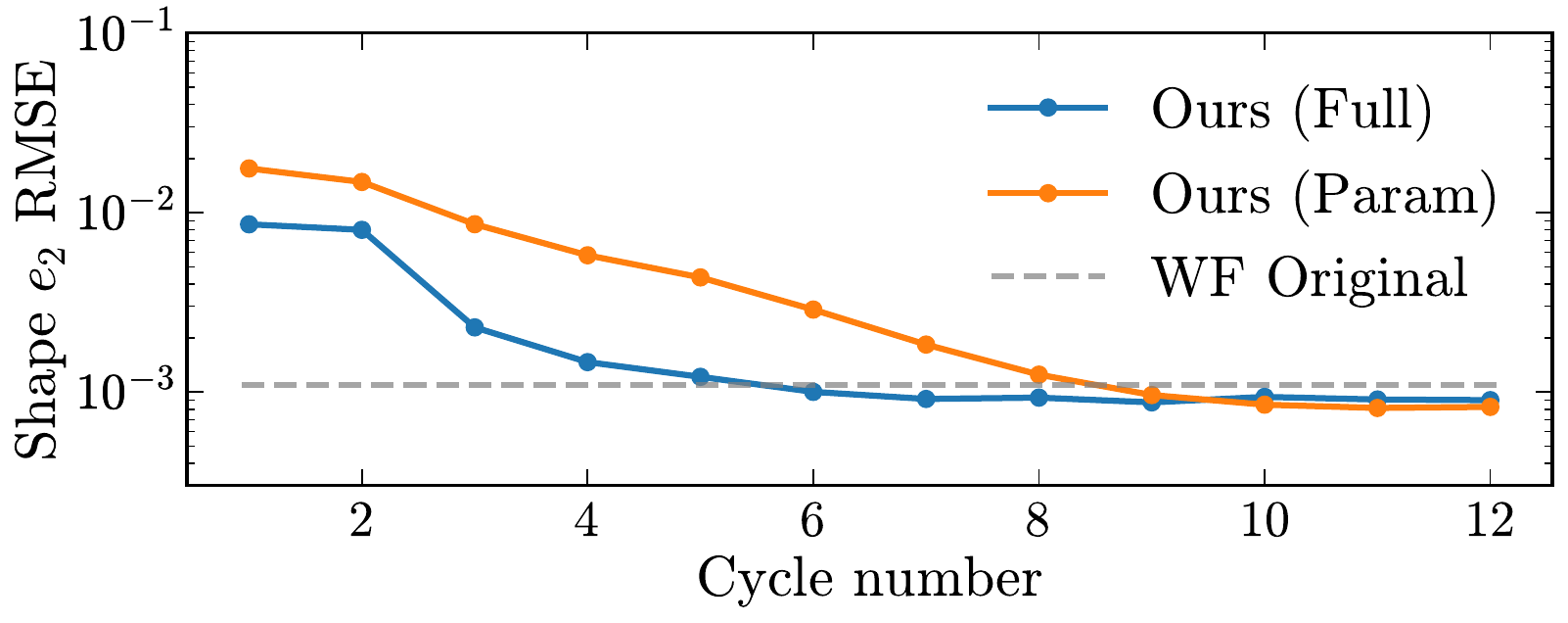}
        \caption{}
        \label{fi:e2_results}
    \end{subfigure}\\
    \begin{subfigure}[b]{\textwidth}
        \centering
        \includegraphics[width=0.4\textwidth]{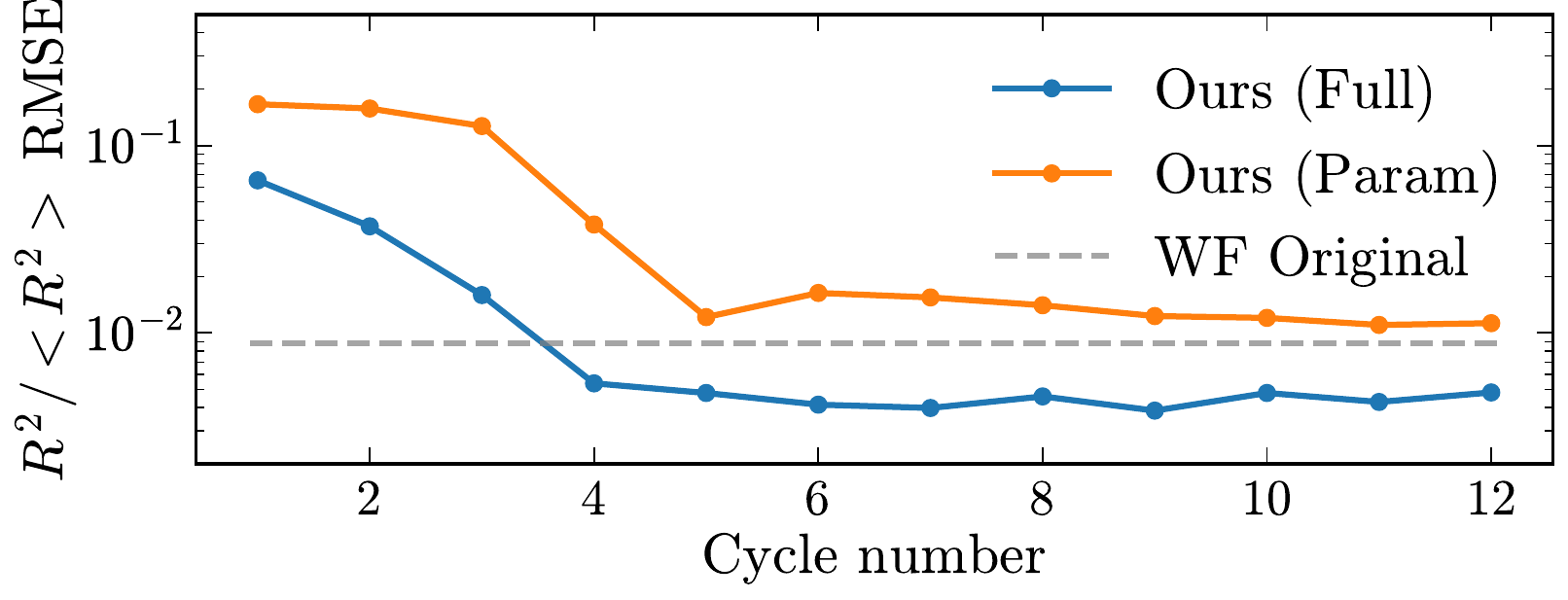}
        \caption{}
        \label{fi:R2_results}
    \end{subfigure}
    \caption{PSF shape results across the 12 optimisation cycles. The "Ours (Full)" model uses both parametric and non-parametric parts for inference, while the "Ours (Param)" model uses only the parametric part. The grey dashed line shows the results of the original optimisation scenario. (a) First ellipticity component: $e_1$. (b) Second ellipticity component: $e_2$. (c) PSF size: $R^2$.}
    \label{fi:shape_results}
\end{figure*}

The ellipticity errors are consistent between the original and new scenarios, although a slight improvement and reduction in uncertainty are observed in the new case when both the parametric and non-parametric models are used for evaluation. 
Given that the shape metrics are computed on the SR pixel PSFs, there is a considerable correlation between the two metrics. 
When compared with the requirements of a Euclid-like survey, the ellipticity error remains about five times higher than the target specification. 
In contrast, the size ($R^2$) error decreases by nearly a factor of two in the new scenario.
The evolution of these metrics over the $12$ optimisation cycles is shown in \autoref{fi:shape_results}. 

\begin{table}
    \centering
    \small
    \caption{Shape results for the WaveDiff model including feature projection and NP reset at the end of each cycle, as in point (iii). The proposed model, "Ours (Full)" uses both parametric and non-parametric parts for inference, while the "Ours (Param)" model uses only the parametric part. The results are compared to the original WaveDiff optimisation strategy presented in \citet{liaudat2023}. The presented metrics are: the two ellipticity components (e1, e2) and the size error, $R^2/\langle R^2\rangle$. For each metric and model, we display the median, standard deviation, and the best of three realisations. The last row shows, for reference, the PSF shape requirements for an Euclid-like survey.}
    \resizebox{0.5\columnwidth}{!}{%
    \begin{tabular}{cccc}
    \hline
    Model       & \multicolumn{3}{c}{Median $\pm$ Std Dev (best) $\times[10^{-3}]$}                      \\ \cline{2-4} 
     & $e_1$  & $e_2$  & $R^2/\langle R^2\rangle$  \\ \hline
    WF Original & $ 1.4 \pm 1.0$ (1.3)  & $1.1  \pm 0.9 $ (1.1) & $8.8 \pm 2.1 $ (7.1) \\
    Ours (Full)    & $1.0 \pm 0.7$ (1.0)   & $1.1 \pm 0.6 $ (0.9)  & $\bm{4.8 \pm 1.4 }$ (4.7) \\
    Ours (Param)   & $1.6  \pm 1.6 $ (1.5) & $0.8 \pm 0.5$ (0.8)   & $9.8 \pm 3.5 $ (8.9) \\ \hline
    Requirements   & $ 0.2 $ & $ 0.2 $   & $ 1.0 $ \\ \hline
    \end{tabular}
    }
    \label{tab:shape_results}
\end{table}

% \FloatBarrier
\FloatBarrier

\section{Absolute WFE results}
\label{apx:absolute_wfe_errors}

In this appendix we present the absolute WFE estimation errors. \Autoref{tab:opd_results_abs} shows the rms error in nm and in units of $\lambda^{-1}$ (considering the central wavelenght $\lambda=725$ nm) for the original model, the new model using both contributions, and the new model using only the parametric component for inference.

\begin{table}[h]
\centering
    \small
    \caption{Absolute WFE errors in nm and units of $\lambda^{-1}$ (considering the central wavelenght $\lambda=725$ nm) for the original WaveDiff model (WF Original) and the proposed model. The model "Ours (Full)" uses both parametric and non-parametric parts for inference, while the "Ours (Param)" model uses only the parametric part. For each case, the median of three realisations is presented along with its respective standard deviation, and in parentheses the realisation with the lowest error.}
    \resizebox{0.5\columnwidth}{!}{%
    \begin{tabular}{@{}ccc@{}}
    \toprule
    Model        & WFE RMS error {[}nm{]}    & WFE RMS error {[}$\lambda^{-1}${]} \\ \midrule
    WF Original  & $22.9 \pm 10.7 \; (19.7)$ & $0.03 \pm 0.01 \; (0.03)$          \\
    Ours (Full)  & $5.1 \pm 1.5 \; (4.9)$    & $0.007 \pm 0.002 \; (0.007)$       \\
    Ours (Param) & $2.7 \pm 0.8 \; (2.4)$    & $0.004 \pm 0.001 \; (0.003)$       \\ \bottomrule
    \end{tabular}
    }
\label{tab:opd_results_abs}
\end{table}

\end{appendix}
\end{document}